\documentclass[review]{elsarticle}

\usepackage{lineno,hyperref}
\modulolinenumbers[5]

\usepackage{amsmath,amsthm,amssymb,epsfig,bm,amsmath}
 \usepackage{graphicx}
\usepackage{float}
\usepackage{amsmath}
\usepackage{mhchem}
\usepackage{xparse}
\usepackage{adjustbox}
\usepackage{MnSymbol}
\usepackage{mathdots}
\usepackage{makecell}
\usepackage{listings}
\usepackage[printwatermark]{xwatermark}
\usepackage{footnote}
\usepackage[flushleft]{threeparttable}
\usepackage{multirow}
\usepackage{relsize}
\usepackage{lettrine}
\usepackage{courier}
\usepackage{color} 
\definecolor{mygreen}{RGB}{28,172,0} 
\definecolor{mylilas}{RGB}{170,55,241}
\usepackage{amsmath,amsthm,amssymb,amsfonts} 
\usepackage{graphicx} 
\usepackage[margin=1in]{geometry} 
\usepackage[yyyymmdd]{datetime}
\usepackage{algorithm}
\usepackage[noend]{algpseudocode}
\usepackage{lipsum}
\usepackage{setspace}
\usepackage{siunitx}
\usepackage{pdfpages}
\usepackage{booktabs,caption}
\usepackage[]{hyperref}
\hypersetup{
  colorlinks   = true, 
  urlcolor     = blue, 
  linkcolor    = black, 
  citecolor   = blue 
}
\usepackage[utf8]{inputenc}

\usepackage{listings}
\usepackage{color}
\definecolor{codegreen}{rgb}{0,0.6,0}
\definecolor{codegray}{rgb}{0.5,0.5,0.5}
\definecolor{codepurple}{rgb}{0.58,0,0.82}
\definecolor{backcolour}{rgb}{0.95,0.95,0.92}
\lstdefinestyle{mystyle}{
    backgroundcolor=\color{backcolour},   
    commentstyle=\color{codegreen},
    keywordstyle=\color{magenta},
    numberstyle=\tiny\color{codegray},
    stringstyle=\color{codepurple},
    basicstyle=\footnotesize,
    breakatwhitespace=false,         
    breaklines=true,                 
    captionpos=b,                    
    keepspaces=true,                 
    numbers=left,                    
    numbersep=5pt,                  
    showspaces=false,                
    showstringspaces=false,
    showtabs=false,                  
    tabsize=2,
    escapeinside={<@}{@>},
}
 
\usepackage{booktabs} 
\usepackage{siunitx} 
\usepackage{bbold}

\usepackage{mathtools} 
\usepackage{gensymb} 
\usepackage{mhchem} 
\usepackage{fixltx2e} 
\usepackage{bbm}

\usepackage{multirow}

\usepackage{fancyhdr} 
\theoremstyle{definition}

\theoremstyle{definition}

\theoremstyle{remark}

\usepackage{soul} 
\soulregister\cite7
\soulregister\ref7
\soulregister\pageref7
\usepackage{bm}

\usepackage{framed} 

\usepackage{multicol} 

\usepackage{nomencl} 
\usepackage{tikz}
\usepackage{wrapfig}
\makenomenclature
\usepackage[resetlabels,labeled]{multibib}

\renewcommand*\nompreamble{\begin{multicols}{2}}

\renewcommand*\nompostamble{\end{multicols}}

\usepackage{amssymb,amsfonts}
\usepackage{pifont}
\definecolor{light-gray}{gray}{0.95}

\def\R{{\mathbb R}}

\def\cU{{\cal U}}
\def\cJ{{\cal J}}
\def\cD{{\cal D}}

\DeclareMathOperator*{\argmin}{arg\,min}

\journal{a Journal (Review in Progress)}

\begin{document}


\begin{frontmatter}

\title{\large Model Calibration of the Liquid Mercury Spallation Target using Evolutionary Neural Networks and Sparse Polynomial Expansions}

\author{Majdi I. Radaideh$^*$, Hoang Tran, Lianshan Lin, Hao Jiang, Drew Winder, Sarma Gorti, Guannan Zhang, Justin Mach, Sarah Cousineau}

\cortext[mycorrespondingauthor]{Corresponding Author: Majdi I. Radaideh (radaidehmi@ornl.gov)}

\address{Spallation Neutron Source, Oak Ridge National Laboratory, 8600 Spallation Dr, Oak Ridge, TN 37830}

\begin{abstract}
\small

The mercury constitutive model predicting the strain and stress in the target vessel plays a central role in improving the lifetime prediction and future target designs of the mercury targets at the Spallation Neutron Source (SNS). We leverage the experiment strain data collected over multiple years to improve the mercury constitutive model through a combination of large scale simulations of the target behavior and the use of machine learning tools for parameter estimation. We present two interdisciplinary approaches for surrogate-based model calibration of expensive simulations using evolutionary neural networks and sparse polynomial expansions. The experiments and results of the two methods show a very good agreement for the solid mechanics simulation of the mercury spallation target. The proposed methods are used to calibrate the tensile cutoff threshold, mercury density, and mercury speed of sound during intense proton pulse experiments. Using strain experimental data from the mercury target sensors, the newly calibrated simulations achieve 7\% average improvement on the signal prediction accuracy and 8\% reduction in mean absolute error compared to previously reported reference parameters. Some individual sensors experienced up to 30\% improvement in their signal prediction accuracy using the calibrated parameters. The proposed calibrated simulations can significantly aid in fatigue analysis to estimate the mercury target lifetime and integrity, which reduces abrupt target failure and saves tremendous amount of costs. However, an important conclusion from this work points out to a deficiency in the current constitutive model based on the equation of state in capturing the full physics of the spallation reaction. Given that some of calibrated parameters that show a good agreement with the experimental data can be nonphysical mercury properties, we need a more advanced two-phase flow model to capture bubble dynamics and mercury cavitation.

\end{abstract}



\end{frontmatter}

\setstretch{1.4}

\section{Introduction}
\label{sec:intro}

The availability of the first target station at the Spallation Neutron Source (SNS) is mission-critical to providing a world-class neutron science program at Oak Ridge National Laboratory (ORNL), supporting a wide range of fundamental scientific studies \cite{mcmanamy2008overview}. Inside the mercury target module, a core component, accelerated high-energy protons strike liquid mercury flowing through a stainless-steel structure to generate pulses of neutrons that are moderated and guided to scientific instruments. The target module receives 60 1-GeV proton beam pulses each second. Each pulse is approximately 0.7 microseconds long and carries 23.3 kJ of energy when the facility operates at 1.4 MW. Approximately, 60\% of the energy from the proton pulses is deposited into the mercury material and the steel target vessel as heat. The energy deposition leads to the creation of a pressure wave, which reacts with the target module structure and creates areas of negative pressure that result in cavitation of the mercury. Simulating the response of the target module structure to the proton pulses represents a significant challenge \cite{riemer2005benchmarking}. Pitting damage on the target’s inner vessel wall, caused by cavitation within the mercury, has been noticed for years \cite{kikuchi2003r,futakawa2010cavitation,mcclintock2012initial}. To maintain the structural integrity of the vessel and extend its life as long as possible, noncondensable helium gas is intentionally injected into the flowing mercury to mitigate cavitation damage \cite{kogawa2015development}. 

Since the response of the stainless-steel target vessel is closely coupled with the mercury that flows through it, an accurate mercury constitutive model (stress-strain relation) is required to predict the strain and stress in the target vessel. The current modeling approach was developed by Riemer \cite{riemer2005benchmarking}. The Riemer model based on the equation of state (EOS) uses a constitutive model for mercury without injected helium gas based on scaled benchmark experiments at LANSCE (Los Alamos Neutron Science Center). The model has two major components. The first component is the EOS for the liquid mercury which connects the volumetric strain to the pressure in the fluid. Secondly, a tensile cutoff (at a certain tensile pressure) is imposed, at which the behavior alters and allows additional volumetric strain without any change in pressure in an attempt to model cavitation. This feature is analogous to a perfectly plastic model. Similar methods have been used to simulate other dynamic fluid-structure interaction problems \cite{espinosa2006novel}. The Reimer model behaves so that if the volumetric strain returns to a smaller value, at which the pressure falls below the tensile cutoff, then the material returns to the standard EOS. The Riemer EOS mercury model has been playing a critical role in the operational prediction of the mercury targets at the SNS. In order to improve the lifetime prediction of current target designs, increase the lifetime of future target designs and minimize disruption to the SNS, accurate simulation of the target behavior via the mercury constitutive model is indispensable. In this paper, we tune the Riemer mercury model by varying its physical model parameters in appropriate ranges through the use of machine learning in order to: 1) inversely search for the optimum parameters, therefore to calibrate the Sierra/Solid Mechanics simulation \cite{team2011sierra} with the machine learning methods;  2) find the possibility to improve the Riemer model \cite{riemer2005benchmarking}; and 3) demonstrate this tuning approach before using it for novel material models intended to represent the mercury with injected helium gas. 

The applications of statistical and machine learning algorithms in this scientific area are indeed limited but continuously growing. Rasheed et al. \cite{rasheed2020deep} employed deep convolutional neural networks (CNN) for the detection of bubble size and bubble distribution in the mercury target. The intelligent detection system proved to be more accurate and efficient in image feature extraction than traditional methods including Circle Hough Transform. Additional study by \cite{garcia2019learning} highlighted the novelty of using machine learning pipeline to analyze neutron scattering data to decipher the structure of materials. A 1D CNN model was developed for classification, followed by random forest regressor to predict cell parameters. The results indicate that their system is much faster and more efficient than traditional methods which are computationally expensive \cite{garcia2019learning}. The model accuracy can reach up to 90\% for shallow models, and for deeper networks, the accuracy can be higher. The use of principal component analysis with an artificial neural network to predict neutron scattering cross-sections was highlighted by \cite{twyman2019machine}. An unsupervised machine learning approach to study phase transitions in single crystal x-ray diffraction data was done by \cite{venderley2019unsupervised}, while a machine learning-based approach to classify the local chemical environment of specific metal families was developed by \cite{lu2019using}. The paper by \cite{doucet2020machine} provides an overview of machine learning activities for neutron scattering at ORNL, which hosts the SNS. Similarly, machine learning applications for neutron and x-ray scattering and spectroscopy research were highlighted by \cite{chen2021machine}. Lastly, the recent review paper by \cite{boehnlein2021artificial} highlights the machine learning research on the broader field of nuclear physics, which includes a section on surrogate modeling with machine learning algorithms. 

Since dealing with the computational cost of the target simulation is cumbersome, the previous model tuning efforts that focus on one-at-a-time adjustments of the simulation parameters \cite{lin2021sensitivity,lin2019tunable} were not effective according to the authors. This is because the process can be tedious once more experiments become available and once the model features more parameters to tune. Therefore, in this work, we propose two approaches to facilitate the model calibration of the expensive simulations of the mercury spallation target. Leveraging the power of machine learning and the measured target strain data, we develop a machine learning framework to minimize the discrepancy between simulations and experimental strain data. The two methods are:
\begin{itemize}
    \item Evolutionary Neural Calibration (ENC): which is summarized by \textit{making a surrogate model of the ``simulation'' and optimizing the surrogate to match the experimental data}. We train and validate feedforward neural network surrogate models for the solid mechanics simulations. A group of evolutionary gradient-free optimization algorithms is used to optimize the surrogate model to match the experimental data. The optimization algorithms are evolution strategies, grey wolf optimizer, differential evolution, and moth-flame optimization. Using an external objective function, these optimization algorithms are executed using different initial guess to ensure a good coverage of the search space.   
    \item Sparse Polynomial Expansions (SPE): which is summarized by \textit{making a surrogate model of the ``discrepancy'' between the simulation and the experiment and optimizing the surrogate to minimize this discrepancy}. A polynomial-based surrogate model is combined with a blackbox optimization using directional Gaussian smoothing (DGS) technique. The surrogate is used to predict the discrepancy between the simulation and the experiment which is a scalar, while the DGS method performs long-range exploration to optimize the polynomial coefficients. Different polynomial spaces for the surrogates are tested to ensure a robust calibration and good coverage of the search space. 
\end{itemize}
Our findings in this work revealed that it is possible to find simulation parameters that can improve the simulation accuracy against experimental data, but they could be nonphysical values for liquid mercury (e.g. density, speed of sound). This indicates that the current physical model based on the equation of state cannot capture the full physics in the mercury target, justifying the need for more advanced two-phase flow model to capture bubble dynamics. 

While each method has its pros and cons, interestingly, the reader will notice a strong agreement between the results of the two methods despite their methodological differences, and a considerable improvement in simulation accuracy compared to experimental data. Furthermore, the two methods feature a general concept, making them suitable for multidisciplinary applications. For the remaining sections of this work, the simulation and experimental setups used to generate the data are described in Section \ref{sec:sns}. In Section \ref{sec:method}, we present the two approaches to solve the model calibration problem: the evolutionary neural calibration and the sparse polynomial expansions. The results of this work are presented and discussed in Section \ref{sec:res}, while the conclusions are highlighted in Section \ref{sec:conc}. 

\section{Neutron Target Simulation and Data}
\label{sec:sns}

\begin{figure}[!h]
    \centering
    \includegraphics[width=0.65\textwidth]{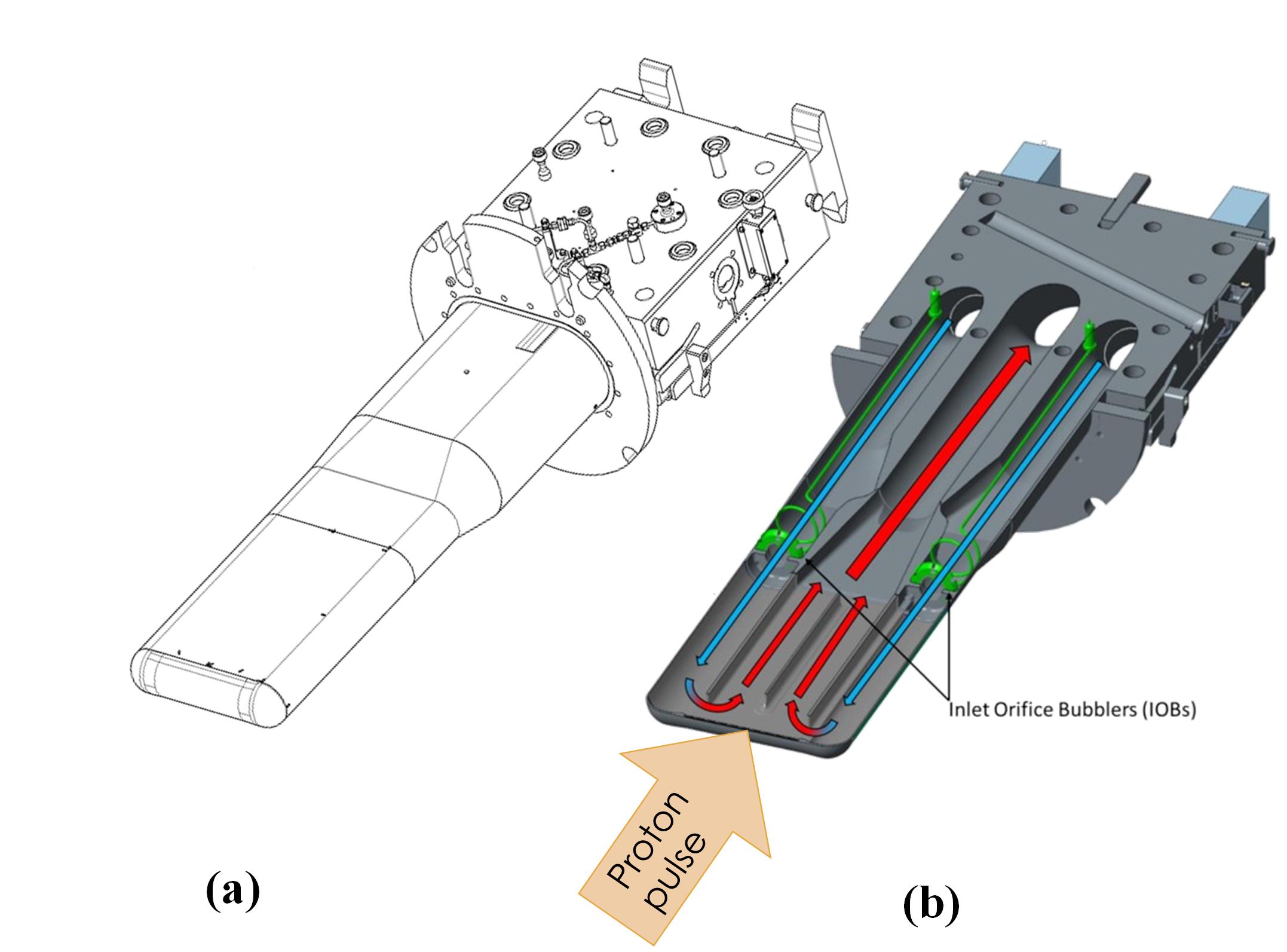}
    \caption{(a) Geometrical sketch of the Mercury target, (b) Mercury flow path within the target}
    \label{fig:target}
\end{figure}

The target module (as shown in Figure \ref{fig:target}) in the first target station at the SNS is subjected to short ($\sim$1 $\mu$s) but intense loading from repeated (60 times per second) proton pulses. The arrows in Figure \ref{fig:target} illustrate the flow direction of mercury within channels of the steel target vessel, the blue color indicates the cool mercury from inlets, and the red color indicates heated mercury flow to the outlets.

\subsection{Experimental Data}
\label{sec:data}
\begin{figure}[!h]
    \centering
    \includegraphics[width=0.8\textwidth]{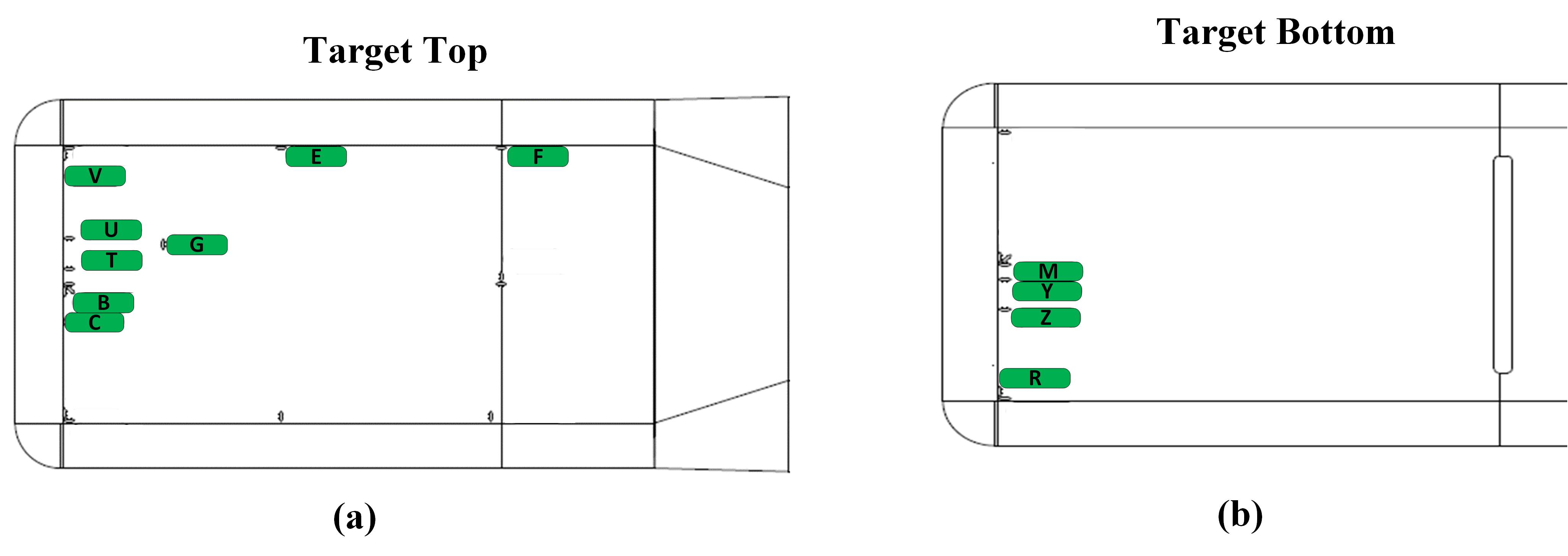}
    \caption{Location of the sensors on the target vessel (a) top and (b) bottom}
    \label{fig:sensors}
\end{figure}

The high energy proton pulse strikes the front of the mercury target material and the stainless-steel target structure, leading to a high-pressure region in both the stationary target structure and the flowing mercury. The suddenly induced pressure wave propagates and leads to fluid-structure interaction. The dynamic deformation of the target vessel can be measured by the strain sensors attached on exterior surface \cite{liu2018strain}. Figure \ref{fig:sensors} highlights the location and label of 12 strain sensors in green blocks, on both the vessel top and bottom surfaces. The small arrows nearby the sensor labels align with the measured strain direction, which is parallel to, perpendicular to, or at 45 degrees to the pulse direction. Within a very short test time, multiple strain data could still be recorded by the sensors due to the high frequency ($\sim$60 Hz) of pulses. Table \ref{tab:sensor_samples} collects all the sensors and their sample number in this study. Note that all these sensor strain data are from pulses equivalent to operating at a 1.4 MW proton power level, without helium gas injection in mercury flow, and on the the specific target design named ``jet flow target'' \cite{barbier2015numerical}. Figure \ref{fig:calib} plots out the strain mean curves and their 95\% confidence interval (CI) bounds by using all the strain data collected in Table \ref{tab:sensor_samples} over 1 millisecond, which is an appropriate time interval during which most of the available sensors experience strain peak from single pulse event. 

\begin{table}[htbp]
  \centering
  \small
  \caption{Number of measured strain data for all sensors in this study}
    \begin{tabular}{ll}
    \toprule
    Sensor & Total Samples \\
    \midrule
    B     & 20 \\
    C     & 40 \\
    E     & 25 \\
    F     & 20 \\
    G     & 20 \\
    M     & 20 \\
    R     & 30 \\
    T     & 25 \\
    U     & 15 \\
    V     & 15 \\
    Y     & 15 \\
    Z     & 15 \\
    \bottomrule
    \end{tabular}%
  \label{tab:sensor_samples}%
\end{table}%


\subsection{Target Simulation}
\label{sec:sim}

\begin{figure}[!h]
    \centering
    \includegraphics[width=0.6\textwidth]{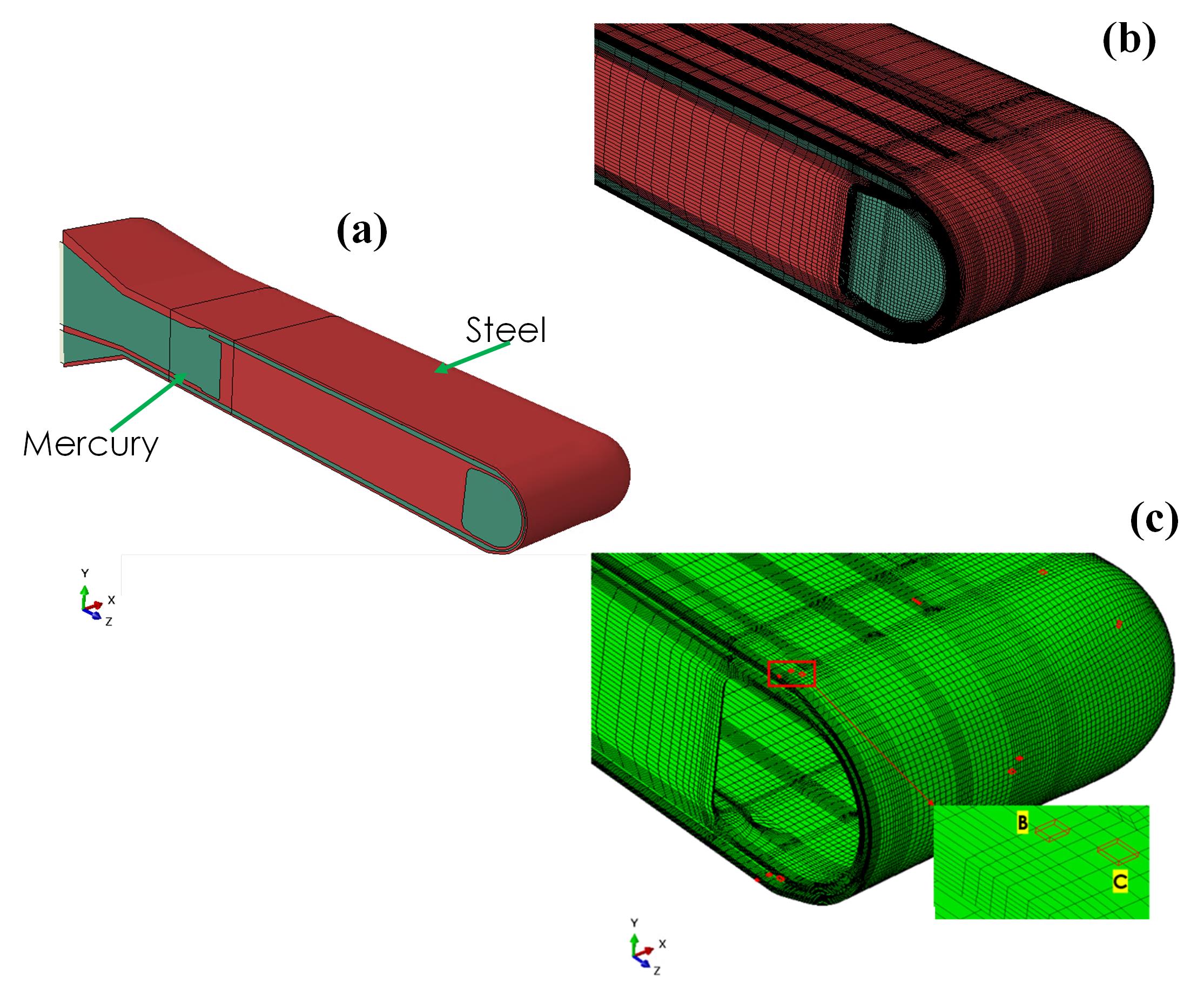}
    \caption{(a) Modeling of the Mercury target in Sierra/Solid Mechanics, (b) Target meshing, (c) Sensor modeling and locations}
    \label{fig:target2}
\end{figure}

Traditional prediction of stress/strain on the target vessel relies on the finite element dynamic simulation, which is based on the material models and parameters introduced by Riemer \cite{riemer2005benchmarking}. The target model is built from real geometry, which contains steel parts and mercury parts, see Figure \ref{fig:target2}(a). The steel part employs an elastic material model, while the mercury part uses the Riemer material model as described in \cite{riemer2005benchmarking}. The energy from a single proton pulse is converted into an initial pressure in the simulation. That spatially-varying initial pressure field triggers response of the mercury and structure from time zero to 1 millisecond. The full Jet-Flow target model (half of real module due to its left-right symmetry) includes 3,971,356 reduced-integration hexahedral elements and 3,650,454 nodes in total, see Figure \ref{fig:target2}(b), with a non-reflecting boundary condition at the back end of mercury surfaces to represent the extended pipes by minimizing any pressure wave reflection from the simulation boundary. The 1 millisecond explicit simulation is performed by Sierra/SolidMechanics \cite{team2011sierra} (Sierra for convenience) 4.56 with the Riemer material model implemented in user subroutine VUMAT, which takes about 25 hours on 96 processors (Intel(R) Xeon(R) E-2174G CPU @ 3.60GHz) with nominal mercury parameters (density = 13,500 kg/m\textsuperscript{3}, sound speed = 1,456 m/s, tensile cutoff = $1.5 \times 10^5$ Pa). Previous simulation tests have demonstrated that stress and strain results from Sierra were identical to the results from ABAQUS/Explicit by using the same hexahedral mesh and material parameters. As shown in Figure \ref{fig:target2}(c), the sensor elements are outlined in red, and they are identified by matching their locations with real strain sensors in pulse test for collecting strain data from simulations. Examples of sensors B and C locations are shown in Figure \ref{fig:target2}(c).

Current Riemer mercury material model uses the above key parameters that were reported to produce a best overall agreement between experimental measurement and finite element simulation \cite{riemer2005benchmarking} for tests without helium gas injection. The tensile cutoff threshold ($\sigma_c$) represents the loss of stiffness at the onset of cavitation in the liquid mercury. The density ($\rho_M$) and sound speed ($c_s$) values are nominal ones for pure mercury material, that could vary a little bit when the mercury gets heated and cavitation occurs in real world. In this work, we tune the Riemer mercury model by varying these three parameters in appropriate ranges using the two proposed methods to find an optimal set of values. Ranges for tensile cutoff threshold, density and sound speed are from 0 to 1.5 MPa, 1350 to 13600 kg/m\textsuperscript{3}, and from 0 to 10000 m/s respectively. The upper bound of mercury density is not far away from its nominal value, by taking the assumption that it has less chance to become very dense. 

\section{Methodology}
\label{sec:method}

The objective of this study is to solve the inverse problem that identifies the optimal values of three simulation settings to match the simulation output to the measured strains.  The three simulation settings are: tensile cutoff threshold ($\sigma_c$), mercury speed of sound ($c_s$), and the initial mercury density ($\rho_M$), where $n_x=3$ is the total number of the calibrated parameters. The problem can be formulated as follows: find the best value of the model input $\vec{x}=[\sigma_c, \rho_M, c_s]$ such that
\begin{equation}
\label{eq:inverse}
    \underset {\vec{x}}{\text{min }} \delta(\vec{x}) = F(\bm{D}, f(\vec{x}))
\end{equation}
where $\delta$ represents the discrepancy to be minimized between the measured data $\bm{D}$ (e.g. from sensors) and the model $f(\vec{x})$, which is the solid mechanics model Sierra. In the next two subsections, we propose two different methods to solve this problem. 


\subsection{Evolutionary Neural Calibration (ENC)}

\begin{figure}[!h]
    \centering
    \includegraphics[width=0.8\textwidth]{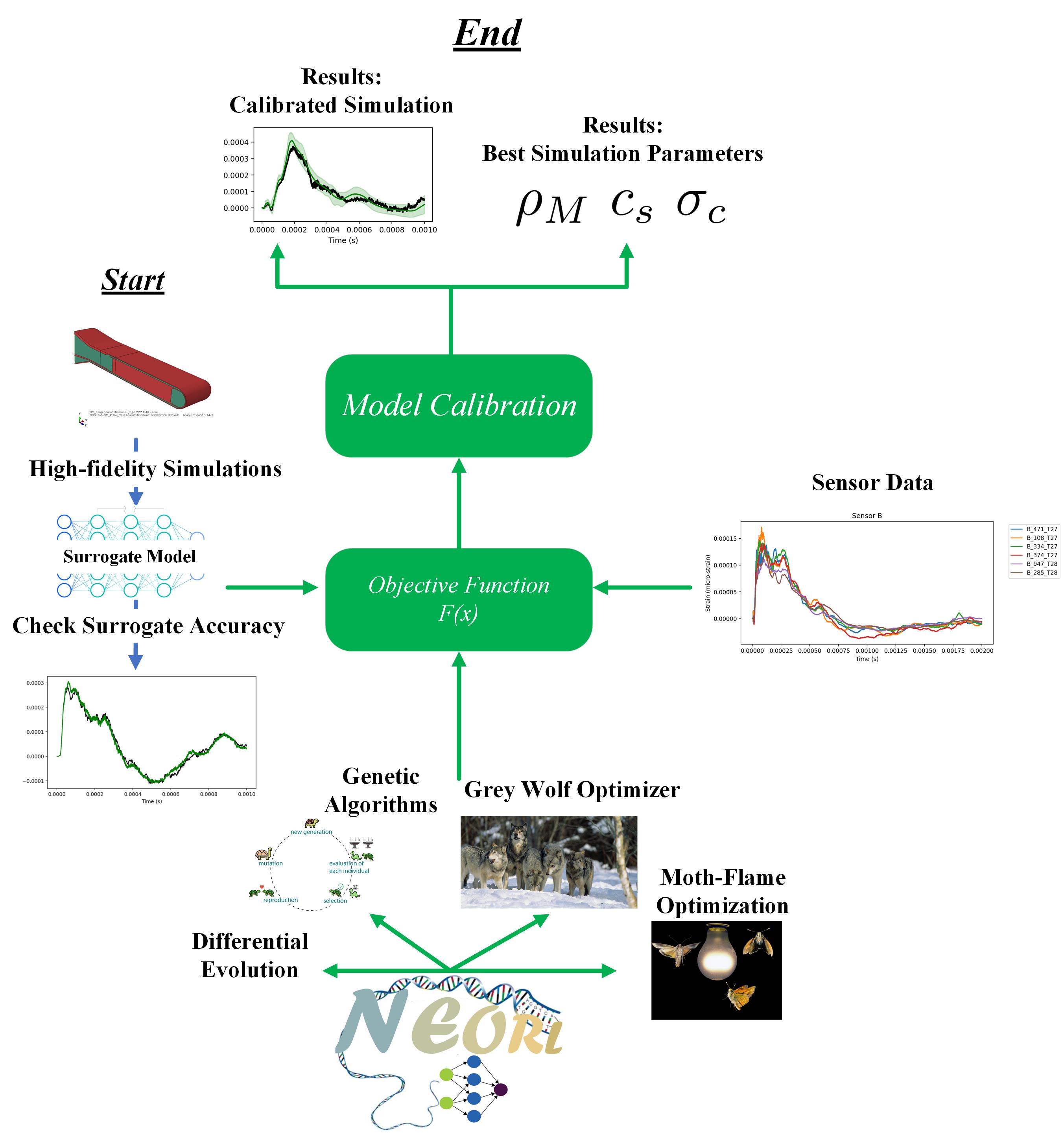}
    \caption{Flowchart of the Evolutionary Neural Calibration (ENC) methodology}
    \label{fig:enc}
\end{figure}

The flowchart of the ENC methodology is shown in Figure \ref{fig:enc}. The method can be described in four major steps:
\begin{enumerate}
    \item Sensor measured data are curated to determine their mean and confidence interval as described in section \ref{sec:data}.
    \item The neural network surrogate model is built and verified from the high-fidelity simulations.
    \item Gradient-free evolutionary optimization algorithms are used to search for the optimal simulation parameters (i.e. $\sigma_c, \rho_M, c_s$).
    \item Items 1-3 are connected in an objective function (i.e. discrepancy) to perform model calibration, so we can solve the inverse optimization problem in Eq.\eqref{eq:inverse}.
\end{enumerate}

In the next subsections, we describe steps 2-4, as step 1 is covered in section \ref{sec:data}.

\subsubsection{Surrogate Modeling}
\label{sec:fnn}

A feedforward neural network (FNN) surrogate model is developed to replace the expensive Sierra computational model. FNN is trained using the classical Tensorflow/Keras framework in Python \cite{chollet2015keras}. The neural network model can be expressed as
\begin{equation}
    \bm{y} = FNN(\vec{x}, \theta)
\end{equation}
where $\theta = (w,b)$ are the network weight and bias parameters to be determined, $\vec{x} =[\sigma_c, \rho_M, c_s]$ is the vector of the three simulation parameters to be calibrated, and $\bm{y}$ is the matrix of strain prediction for all time steps and all sensors as shown for the experimental data in Figure \ref{fig:calib}. Therefore, the output of the surrogate in ENC is a two-dimensional array of time $\times$ sensor. The FNN consists of an input layer, multiple hidden layers, and an output layer. The output of each hidden layer can be calculated as

\begin{equation}
    z=\sum_{i=1}^nw_ix_i+b,
\end{equation}
where the weighted inputs ($wx$) are summed together for all nodes ($n$) in the hidden layer, and a constant value called bias ($b$) is added to produce the layer input. Afterward, the layer input is passed to an activation function ($g$) to produce the layer output 

\begin{equation}
    y=g(z)=g\Big(\sum_{i=1}^nw_ix_i+b\Big).
\end{equation}

The activation function is a non-linear mapping function to model the relationship between the input and the output. There are many common options for the activation function. Rectified Linear Unit (ReLU) is a typical activation function for FNN
\begin{equation}
    g(z)=max(0,z),
\end{equation}
or the sigmoid function, which can be expressed as 
\begin{equation}
    g(z)=\frac{1}{1+e^{-z}},
\end{equation}
which has a continuous range between [0,1]. Other activation functions are leaky ReLU, hyperbolic tangent (tanh), and linear functions. This process can be generalized for $k$ hidden layers as 
\begin{equation}
    \bm{h}_k=g_i(\bm{w}_k\bm{h}_{k-1}+\bm{b}_k),
\end{equation}
where $g_k(.)$ is the activation function for hidden layer $k$ and $\bm{h_{k-1}}$ is the output of the previous hidden layer. After completing the forward propagation step in all layers, the predicted output ($\hat{y}$) can be determined, which allows determining the loss function, such as the mean absolute error (MAE)
\begin{equation}
\label{eq:cost}
    \quad L (\bm{w},\bm{b})= MAE = \frac{1}{n}\sum_{i=1}^n |\hat{y}(\bm{w},\bm{b})_i-y_i|,
\end{equation}
where $y$ is the target output (e.g. from Sierra) and $n$ is the number of training samples. For each training step, the errors in Eq.{\eqref{eq:cost}} are propagated backwards through their gradient values, where the weights ($\bm{w}$) of each layer are updated according to the loss gradient ($\frac{\partial L}{\partial w_i}$), such that the loss function is more likely to be minimized in the next network update. The backpropagation step can be trained by the gradient descent or using more advanced techniques such as the Adam optimizer (adaptive moment estimation). In addition to the number of hidden layers, and the number of nodes per layer, other hyperparameters for the neural network to be determined are:
\begin{itemize}
    \item Adam optimizer has a hyperparameter called learning rate, which controls the step size the optimizer takes while searching for the minimum of the loss function.
    \item Every full forward and backward pass through the dataset is called ``epoch''. The FNN network needs to be trained for many epochs (e.g. 100) to achieve satisfactory accuracy.
    \item The dataset is passed to the FNN in multiple batches with a mini-batch size to be determined (e.g. 8, 16, 32). This practice allows multiple model updates per epoch, and also accommodates the available memory.
    \item The dataset is split into training and test sets to avoid model overfitting to the training data. A typical split of 80\% for training and 20\% for testing is usually used. 
\end{itemize}

\subsubsection{Evolutionary Optimization}
\label{sec:neorl}

Evolutionary and swarm computation is a family of algorithms for optimization inspired by biological evolution. Despite the diverse range of the evolutionary algorithms, their algorithmic structure can be broadly summarised in Algorithm \ref{alg:ea}. These algorithms feature population-based trial and error problem solvers with a metaheuristic theme. An initial set of candidate solutions is generated and iteratively updated using natural operators and guided by their objective function value. In each new generation, the population (with size $pop\_{size}$) is produced by removing less desired solutions, introducing small random changes to mimic search exploration, and/or pruning fit solutions to mimic search exploitation. Each individual is a vector of $n_x$ components, where these components are the parameters to be optimized, i.e. $\vec{x} =[\sigma_c, \rho_M, c_s]$. The core difference between these algorithms originates from the natural operators being used to update the population every generation. 

\begin{algorithm}[!h]
  \small
    \caption{Pseudocode for evolutionary algorithms (Minimization)}
    \label{alg:ea}
    \begin{algorithmic}[1]
     \State \textbullet Set algorithm hyperparameters
     \State \textbullet Define the objective function and the input space bounds.
     \State \textbullet Initialize the population randomly or with a guess with size $pop\_size$, i.e. $\{\vec{x}_1^0, \vec{x}_2^0, ..., \vec{x}_{pop\_size}^0\}$
     \State \textbullet Initialize $y_{best}$=$\infty$
     \For{Generation $k = 1$ to $N_{gen}$} 
             \For{Individual $j = 1$ to $pop\_size$}
                \State \textbullet Individual $\vec{x}_j$ is evaluated by the objective function $\delta(\vec{x})$ in Eq.\eqref{eq:obj}, and the value $y_j$ is returned
                 \If {$y_j < y_{best}$}
                    \State \textbullet  Set $y_{best} = y_j$
                    \State \textbullet  Set $\vec{x}_{best} = \vec{x}_j$
                 \EndIf
            \EndFor 
        \State \textbullet (A special step for ES) Select the top $m$ individuals to survive to the next generation, i.e. $m < pop\_size$ 
        \State \textbullet Update the population position using evolutionary operations, Eq.\eqref{eq:de}-Eq.\eqref{eq:gwo3}.
    \EndFor 
    \State \textbullet  Return $\vec{x}_{best}$ and $y_{best}$
\end{algorithmic}
\end{algorithm}

As evolutionary optimization algorithms (like other algorithms) in general cannot guarantee finding a global optima, in this work, we use a combination of multiple evolutionary algorithms. This improves the exploration of the search space and maximizes the likelihood of finding a robust optimal solution. The algorithms being used here are: evolution strategies (ES)\cite{beyer2002evolution}, grey wolf optimizer (GWO)\cite{mirjalili2014grey}, differential evolution (DE) \cite{storn1997differential}, and moth-flame optimization (MFO) \cite{mirjalili2015moth}.

ES \cite{beyer2002evolution} inherit most of the features of the popular genetic algorithm such as the crossover and mutation operators. In every generation, the population individuals are mixed through the biological crossover of the genes (i.e. $n_x$ inputs) with probability ($cxpb$), and then a mutation is applied to few of these genes with probability ($mutpb$). The major difference between ES and the classical genetic algorithms is the introduction of the strategy vector that adapts and varies the mutation rate during the search for every gene/input in the individual, compared to a fixed mutation rate for all inputs in classical genetic algorithms. 

DE \cite{storn1997differential} updates each individual in the population by randomly drawing three individuals: $\vec{a}, \vec{b}, \vec{c}$, then the new individual position can be expressed as
\begin{equation}
\label{eq:de}
    \vec{x}_i'=\vec{a}_i + RF(\vec{b}_i - \vec{c}_i), \quad \text{if } r_i < CR; \quad r_i \sim \mathcal{U}[0,1],
\end{equation}
otherwise, if $r_i \geq CR$, set  $\vec{x}_i'=\vec{x}_i$, where $\vec{x}_i$ is the current position. Here, $CR \in [0,1]$ is the crossover probability and $RF \in [0,2]$ is the recombination factor, both are tuned parameters. Also, the $i$ index loops through the input parameters, i.e. $i=1,...,n_x$.

The MFO \cite{mirjalili2015moth} defines two matrices called $M$ for moth positions and $F$ for the flames. Both matrices have a size of $pop\_size \times n_x$. Afterward, the position update of the moths can be done using \cite{mirjalili2015moth}
\begin{equation}
\label{eq:mfo}
    M_i = D_i e^{bt} \times cos(2\pi t) + F_j
\end{equation}

where $D_i = |F_j - M_i|$ is the distance between the moth $i$ and the flame $j$, $b$ is a constant defining the shape of the logarithmic spiral, and $t \sim \mathcal{U}[-1,1]$.

GWO \cite{mirjalili2014grey} models the social hierarchy of grey wolves by considering the fittest solution as the alpha wolf ($\alpha$). The second and third best solutions are named
beta ($\beta$) and delta ($\delta$) wolves, respectively. The remaining solutions in the group are named as omega ($\omega$). In the GWO algorithm, the hunting or optimization process is guided by $\alpha$, $\beta$, and $\delta$ wolves, while the $\omega$ wolves follow these three leaders. The hunting of the wolves can be modeled as \cite{mirjalili2014grey}:

\begin{equation}
\label{eq:gwo1}
    \vec{D}_k = |\vec{C}_i . \vec{x}_k-\vec{x}|, \quad k = \alpha, \beta, \delta, \quad i=1, 2, 3,
\end{equation}
\begin{equation}
\label{eq:gwo1}
    \vec{x}_i = \vec{x}_k - \vec{A}_i . \vec{D}_k, \quad k = \alpha, \beta, \delta, \quad i=1, 2 , 3,
\end{equation}
where the updated position of the wolf in the next generation is simply the average of the three calculated positions:
\begin{equation}
\label{eq:gwo3}
    \vec{x}'=\frac{\vec{x}_1 + \vec{x}_2 + \vec{x}_3}{3},
\end{equation}
where the two vectors $\vec{A}$ and $\vec{C}$ are defined as 
\begin{equation}
    \vec{A} = 2 \vec{a}.\vec{r}_1 - \vec{a}, \quad \vec{r}_1 \sim \mathcal{U}[0,1], 
\end{equation}
\begin{equation}
    \vec{C} = 2 \vec{r}_2, \quad \vec{r}_2 \sim \mathcal{U}[0,1],
\end{equation}
where the components of $\vec{a}$ are linearly decreased from 2 to 0 over the total number of generations ($N_{gen}$).

In this work, we execute these algorithms randomly and independently with different initial guesses and we record the best solution achieved by the executed algorithm. We fix the number of generations and population size in every round. For example, in round 1, one of the four algorithms is selected randomly (e.g. GWO) and executed for number of generations. Then, the process is repeated for about 100 rounds, and the best solution is recorded for every round. This randomness and usage of multiple algorithms ensures independence, flexibility, and better coverage of the search space. 
NEORL (NeuroEvolution Optimization with Reinforcement Learning) is a set of implementations of hybrid algorithms combining neural networks and evolutionary computation based on a wide range of machine learning and evolutionary intelligence architectures \cite{radaideh2021neorl, radaideh2021physics}. NEORL has been developed by one of the authors of this current study, and it offers a robust implementation for ES, GWO, MFO, and DE, and so we utilize it in this work. 

\subsubsection{Objective Function}
\label{sec:delta}

The objective function is the component that brings the optimizer, the surrogate model, and the measured data together in this method. As the objective function is always problem-dependent and there could be many ways to formulate it, in this work, our objective function focuses on two main goals. First, \textbf{minimizing} the mean absolute error (MAE) between the simulation and the experiment mean (see Figure \ref{fig:calib}). Second, \textbf{maximizing} the simulation accuracy by maximizing the number of simulation time steps that fall within the experiment 95\%-CI (see Figure \ref{fig:calib}). The first objective can be formulated as 

\begin{equation}
\label{eq:f1}
MAE = f_1(\vec{x}) = \frac{1}{N^s}\frac{1}{N^t}\sum_{i=1}^{N^{s}}\sum_{j=1}^{N^{t}} |y_{i,j} - \textbf{D}_{i,j}|,    
\end{equation}
where $N^t$ is the number of the simulation time steps (i.e. 10000), $N^s$ is the number of sensors (i.e. 12), $y_{i,j}$ is the surrogate prediction for time step $j$ and sensor $i$, and $\textbf{D}_{i,j}$ is the corresponding experimental mean value. The second objective is

\begin{equation}
\label{eq:f2}
ACC = f_2(\vec{x}) = \frac{1}{N^s}\frac{1}{N^t}\sum_{i=1}^{N^{s}}\sum_{j=1}^{N^{t}} \mathbb{1}_{CI^{L}_{i,j} < y_{i,j} < CI^{U}_{i,j}},     
\end{equation}
where $\mathbb{1}$ is the indicator function, $CI^{L}_{i,j}$ is the experiment lower bound, and $CI^{U}_{i,j}$ is the experiment upper bound for time step $j$ and sensor $i$. Now the final objective can be written as

\begin{equation}
\label{eq:obj}
\vec{x}^* = \underset {\vec{x}}{\text{max }} \delta(\vec{x}) = f_2(\vec{x}) - f_1(\vec{x}).    
\end{equation}

We can notice that this is a multiobjective problem of two objectives, where the difference between the accuracy and the error functions is maximized. The optimal values of the input parameters that achieve this maximum difference is $\vec{x}^*=[\sigma_c, \rho_M, c_s]$. 

\subsection{Sparse Polynomial Expansions (SPE)}

The method can be described in four major steps:
\begin{enumerate}
    \item Curate and analyze the sensor measurement data to determine their mean and confidence interval (described in section \ref{sec:data}).
    \item Define a suitable discrepancy between sensor measurements and simulation data. The discrepancy function should push the simulation toward the experimental mean and at the same time heavily penalize simulation data that lie outside the confidence interval (see section \ref{sec:disc_func}).
    \item Compute the value of discrepancy at sampled parameters, and construct a surrogate of the discrepancy for the entire searching domain using sparse polynomial approach (see section \ref{section:SPE}). 
    \item Search for the optimal parameters using a state-of-the-art optimization with nonlocal gradient, namely, directional Gaussian smoothing (DGS) optimization (see section \ref{section:DGS}). 
\end{enumerate}

\subsubsection{Discrepancy function}
\label{sec:disc_func}
An ideal simulation should generate strain data which fall completely into the confidence interval and is close to the experimental mean. To measure the difference between simulation and experimental mean, we use $L^2$ relative norm accumulated on all sensors: 
\begin{align}
    \label{eq:RRMSE}
    \delta_1(\vec{x}) = \sum_{i=1}^{N^{s}} \frac{ {\|y_{i} - \textbf{D}_{i}\|_2}}{ {\| \textbf{D}_{i}\|_2}} = \sum_{i=1}^{N^{s}} \frac{ \sqrt{\sum_{j=1}^{N^{t}} |y_{i,j} - \textbf{D}_{i,j}|^2}}{ \sqrt{\sum_{j=1}^{N^{t}}| \textbf{D}_{i,j}|^2}}. 
\end{align}
Here, $\|\cdot\|_2$ denotes the $L_2$ vector norm, $y_{i}$ and $\textbf{D}_{i}$ are the surrogate prediction and corresponding experimental mean value for sensor $i$, and $y_{i,j}$ and $\textbf{D}_{i,j}$ are those data for each sensor and time step, as in section \ref{sec:delta}. 

Next, to enforce the simulation to stay inside the confidence interval, we apply the relative max norm on the excess value of simulation data over the experiment upper bound or under the experiment lower bound, accumulated on all sensors:
\begin{align}
    \label{eq:Linf}
    \delta_2(\vec{x}) = \sum_{i=1}^{N^{s}} \max_{1\le j \le N^t} \frac{ {|( y_{i,j} - CI^{U}_{i,j})^+|}}{ {| CI^{U}_{i,j} - CI^{L}_{i,j}|}} +  \sum_{i=1}^{N^{s}} \max_{1\le j \le N^t} \frac{ {|(CI^{L}_{i,j} - y_{i,j})^+|}}{ {| CI^{U}_{i,j} - CI^{L}_{i,j}|}}. 
\end{align}
Here, $CI^{U}_{i,j}$ and $CI^{L}_{i,j}$ are the experiment upper and lower bound, and $(\cdot)^+$ is the positive part function, i.e., $(y)^+ = \max\{y,0\}$. Summing $\delta_1$ and $\delta_2$, with a weight applied on $\delta_2$ to bring values of the two functions to the same scale, we obtain the final discrepancy function, which is to be minimized to find the optimal parameters: 
\begin{align}
\label{eq:loss}
\delta(\vec{x}) = \delta_1(\vec{x}) + 5 \delta_2(\vec{x}). 
\end{align}

\subsubsection{Surrogate modeling}
\label{section:SPE}
Let $\vec{x}_1,\ldots,\vec{x}_n$ be $n$ sampled parameters for training the surrogate. For each parameter $\vec{x}_k$ ($1\le k\le n$), we acquire the simulation data $f(\vec{x}_k)$, from which the discrepancy $\delta(\vec{x}_k) = F(\bm D,f(\vec{x_k}))$ will be computed. We proceed to construct a surrogate $\widetilde{\delta}$ of $\delta$ in the entire searching domain. We adopt a polynomial approximation approach based on compressed sensing, also known as sparse polynomial approximation, due to its simplicity, low cost and data efficiency. In particular, theory of compressed sensing indicates that if a function possesses a sparse representation in some basis, it can be reconstructed from a limited number of sampling points, only scaling linearly with the sparsity and \textit{independent} of the size of representation \cite{FouRau13,RauWard12,Adcock15b,ChkifaDexterTranWebster15,TranWebster_NMPDE21}. This sampling requirement is much milder than those by traditional methods such as Galerkin projection and interpolation. 

Let the discrepancy function $\delta: \cU\to  \mathbb{R}$ be defined on the searching domain $\cU\subset \mathbb{R}^{n_x}$ ($n_x=3$), endowed with a probability measure $\varrho$. We represent $\delta$ by the finite expansion
\begin{align}
\label{leg_exp}
\delta(\vec{x}) \simeq \sum_{{j} \in \cJ} c_{j} \Psi_{j}(\vec{x}),\ \ \vec{x}\in \cU,
\end{align}
where $\{\Psi_j\}_{j\in \cJ}$ is a pre-determined orthonormal system associated with $\varrho$ and indexed by a finite set $\cJ$. We define the corresponding polynomial subspace $\mathbb{P}_{\cJ} := \text{span}\{\Psi_j(\vec{x}): j\in \cJ\}$. In compressed sensing approach, we aim to recover $\delta$ by reconstructing unknowns $\{c_{j}\}_{{j}\in \cJ}$ from $n$ samples $\delta(\vec{x}_1),$ $\ldots, \delta(\vec{x}_n)$, which are drawn independently from probability measure $\varrho$. Denote the cardinality of $\cJ$ by $N$, i.e., $N=\#(\cJ)$, the normalized sampling matrix by $\bm A$ and the normalized observation of the target function by $\bm{\delta}$, i.e.,
\begin{align*}
\bm{A} := \left(\frac{\Psi_j({\vec{x}}_i)}{\sqrt{n}}\right)_{\substack{1\leq k\leq n \\ j\in \cJ}}  \in \R^{n\times N} ,\quad \text{and}\quad
\bm{\delta} := \left(\frac{\delta({\vec{y}}_i)}{\sqrt{n}}\right)_{1\le k \le n} \in \mathbb{R}^n, \end{align*}
this task amounts to finding for the coefficient vector $\bm c = (c_{j})_{j\in \cJ}$ that satisfies $\bm A  \bm c \simeq \bm g$. Since the simulations $\delta(\vec{x}_k)$ are costly to acquire, we want to reconstruct $\bm c$ with fewest possible samples. Although the problem becomes underdetermined in that case, this is feasible assuming that the expansion Eq.\eqref{leg_exp} is sparse, in which case $\bm{c}$ can be reconstructed by the $\ell_1$-regularized problem: 
\begin{align}
\label{intro:l1}
\min_{\bm z\in \mathbb{R}^N}  \left(\| \bm A \bm z - \bm \delta\|^2_2 + \lambda \|\bm z\|_1\right), 
\end{align}
where $\lambda$ is the regularization parameter and $\|.\|_2$ and $\|.\|_1$ are the $L^2$ and $L^1$ vector norms, respectively. In this paper, we select Chebyshev polynomial bases for $\{\Psi_j\}_{j\in \cJ}$, due to its superior sampling efficiency compared to other orthonormal systems \cite{FouRau13,ChkifaDexterTranWebster15}, therefore, $\varrho$ is the Chebyshev probability measure. 
The set of simulations is split into training and validation set (80\% for training and 20\% for validation). We solve the non-smooth optimization problem Eq.\eqref{intro:l1} with training data using CVXPY package \cite{diamond2016cvxpy,agrawal2018rewriting}. 

\subsubsection{Directional Gaussian smoothing (DGS) optimization}
\label{section:DGS}
After the surrogate modeling step, we obtain an accurate approximation $\widetilde{\delta}$ of the discrepancy function $\delta$. We can inspect and optimize $\delta$ by inquiring $\widetilde{\delta}$, without the need of additional simulations. To find the optimal parameters for the Riemer material model ($\vec{x}^*=[\sigma_c, \rho_M, c_s]$), we solve the optimization problem 
\begin{align}
\label{eq:optim}
    \vec{x}^* = \argmin_{\vec{x}\in \cU} \widetilde{\delta}({\vec{x}}). 
\end{align}
Preliminary inspection of $\widetilde{\delta}$ shows that the objective function is multimodal with multiple local minima (see Section \ref{sec:SPE_results}), so we adopt the DGS optimization approach. This method was recently developed in \cite{DGS_UAI,AdaDGS21} and applied to reinforcement learning and scientific problems in \cite{1935-9179_2021_6_4119,ZHANG2021109213}, where it has been demonstrated very efficient for this type of functions. DGS has different concept than the population-based optimization methods used by ENC, which gives SPE another methodological difference.  

\begin{wraptable}{r}{0.52\textwidth}
\begin{tabular}{p{0.48\textwidth}}
\toprule
\vspace{-0.4cm} 
{{\bf Algorithm 2}: Adaptive DGS algorithm} \\
\midrule
\vspace{-0.3cm}
\begin{algorithmic}[1]
    \State {\bf Hyper-parameters}: $M$: number of GH quadrature points, $\sigma_0$: initial smoothing radius, $\gamma$: tolerance for resetting the smoothing radius.
    \State {\bf Input:} The initial state $\vec{x}^{(0)}$.
    \State {\bf Output:} The final state $\vec{x}^{(T)}$.
    \For {$i=0, \ldots, T-1$}
        \State Evaluate $\{\widetilde{\delta}(\vec{x}^{(i)} + \sqrt{2}\sigma_i v_m \vec{\xi}_{k})\}^{k=1, \ldots, d}_{m =1, \ldots, M}$
        \For{$k = 1, \ldots, d$}
            \State Compute $\cD^M [\widetilde{\delta}_{\sigma_i}^{\, \vec{\xi}_k}(\vec{x}^{(i)})]$ as in Eq.~(\ref{e8})
        \EndFor
        \State Assemble $ {\nabla}^M_{\sigma_i}[\widetilde{\delta}](\vec{x}^{(i)})  $ as in Eq.~(\ref{e5})
        \State Update $\alpha_i$ using backtracking line search
        \State Set $\vec{x}^{(i+1)} = \vec{x}^{(i)} - \alpha_i {\nabla}^M_{\sigma_i}[\widetilde{\delta}](\vec{x}^{(i)})$ as in Eq.~(\ref{DGS:GD})
        \State Set $\sigma_{i+1} = \frac{1}{2}(\sigma_{i} + \alpha_{i} )$ 
        \If{$|\widetilde{\delta}(\vec{x}^{(i+1)}) - \widetilde{\delta}(\vec{x}^{(i)})|/|\widetilde{\delta}(\vec{x}^{(i)})| < \gamma$}
            \State Set $\sigma_{i+1} = \sigma_0$ 
        \EndIf
    \EndFor
\vspace{-0.4cm}
\end{algorithmic}\\\bottomrule
\end{tabular}
\end{wraptable}

In optimizing a multimodal or noisy function where the gradient information is not available or not useful, one well-known approach is to smooth the objective function with Gaussian smoothing and use the directional derivatives or gradient of the smoothed function to guide the search for optima \cite{NesterovSpokoiny15}. While these search directions are able to achieve long-range exploration and less susceptible to highly fluctuating landscapes, they may be challenging to compute \cite{salimans2017evolution,Berahas_FoCM}. Directional Gaussian smoothing is a new method that can find the nonlocal search direction via Gaussian smoothing in an innovative way. The key idea of DGS is to conduct 1D long-range explorations along $d$ orthogonal directions in $R^d$, each of which defines a non-local directional derivative as a 1D integral. The Gauss-Hermite quadrature is then used to estimate the 1D integrals to provide accurate estimation of the DGS gradient. In particular, consider a 1D cross section of $\widetilde{\delta}(\vec{x})$ along direction $\vec{\xi}$, i.e., $\widetilde{\delta}(\vec{x} + v\vec{\xi})$. The Gaussian smoothing of $\widetilde{\delta}(\vec{x})$ along $\vec{\xi}$ is  
\begin{align*}
    \widetilde{\delta}_\sigma^{\, \vec{\xi}}(\vec{x}) := \frac{1}{\sqrt{2\pi}} \int_{\mathbb{R}} \widetilde{\delta}(\vec{x} + \sigma v\vec{\xi}) \exp(-{v^2}/{2}) dv,
\end{align*}
where $\sigma$ is the smoothing radius. The derivative of the smoothed  $\widetilde{\delta}_\sigma^{\, \vec{\xi}}(\vec{x})$ along $\vec{\xi}$ is a 1D expectation
\begin{equation*}
    \cD [\widetilde{\delta}_\sigma^{\, \vec{\xi}}(\vec{x})] 
     = \frac{1}{\sigma\sqrt{2\pi}} \int_{\mathbb{R}} \widetilde{\delta}(\vec{x} + \sigma v\vec{\xi}) \exp(-{v^2}/{2}) v dv,
\end{equation*}
where $\cD[\cdot]$ denotes the differential operator. Intuitively, the DGS gradient is formed by assembling these directional derivatives on $d$ orthogonal directions 
\begin{equation*}
{\nabla}_{\sigma}[\widetilde{\delta}](\vec{x}) := \Big[\cD [\widetilde{\delta}_\sigma^{\, \vec{\xi}_1}(\vec{x})] , \cdots, \cD [\widetilde{\delta}_\sigma^{\, \vec{\xi}_d}(\vec{x})] \Big],
\end{equation*}
where the set $\{\vec{\xi}_1,\ldots, \vec{\xi}_d\}$ forms the standard basis in $\mathbb{R}^d$. Since each component of ${\nabla}_{\sigma}[\widetilde{\delta}](\vec{x})$ only involves a 1D integral, the Gauss-Hermite (GH) quadrature rule can be applied to approximate it with high accuracy
\begin{align}
\label{e8}
 \cD^M [\widetilde{\delta}_\sigma^{\, \vec{\xi}}(\vec{x})] 
   =
      \frac{1}{\sqrt{\pi}\sigma} \sum_{m = 1}^M w_m \,\widetilde{\delta}(\vec{x} + \sqrt{2}\sigma v_m \vec{\xi})\sqrt{2}v_m. 
\end{align}

Here $\{v_m\}_{m=1}^M$ are the roots of the $M$-th order Hermite polynomial and $\{w_m\}_{m=1}^M$ are quadrature weights, the values of which can be found in \citep{Handbook}. Applying the GH quadrature to each component of ${\nabla}_{\sigma}[\widetilde{\delta}](\vec{x})$, the following estimator is defined for the DGS gradient
\begin{equation}
\label{e5}
  {\nabla}^M_{\sigma}[\widetilde{\delta}](\vec{x}) = \Big[\cD^M [\widetilde{\delta}_\sigma^{\, \vec{\xi_1}}(\vec{x})] , \cdots, \cD^M [\widetilde{\delta}_\sigma^{\, \vec{\xi}_d}(\vec{x})] \Big]. 
\end{equation}

Then, the DGS gradient is readily integrated to first-order schemes to replace the local gradient. We consider the gradient descent scheme with DGS
 \begin{equation}
 \label{DGS:GD}
\vec{x}^{(i+1)} = \vec{x}^{(i)} - \alpha_i {\nabla}^M_{\sigma_i}[\widetilde{\delta}](\vec{x}^{(i)}), 
\end{equation}
where $\vec{x}^{(i)}$ and $\vec{x}^{(i+1)}$ are the candidate solutions at iteration $i$ and $i+1$, and $\alpha_i$ is the learning rate. In this work, we employ an adaptive version of DGS, first developed in \cite{AdaDGS21}, which automatically updates the two important hyperparameters, i.e., the smoothing radius $\sigma_i$ and the learning rate $\alpha_i$. Briefly, this algorithm uses backtracking line search to estimate the optimal learning rate, and then uses the acquired step size to update the smoothing radius following the simple rule $\sigma_i = \frac{1}{2}(\sigma_{i-1} + \alpha_{i-1} )$. Occasionally, we reset the smoothing radius if the method stops to make sufficient progress. A pseudo-code of the algorithm is presented in Algorithm {\color{blue} 2}. For more details, we refer to \cite{AdaDGS21}.

\section{Results and Discussion}
\label{sec:res}

The results are presented in three subsections. First, ENC results are presented, followed by the results of the SPE method. Then, the results are validated with the Sierra computer code and discussed. 

\subsection{Evolutionary Neural Calibration Results}

\begin{figure}[!h]
    \centering
    \includegraphics[width=.5\textwidth]{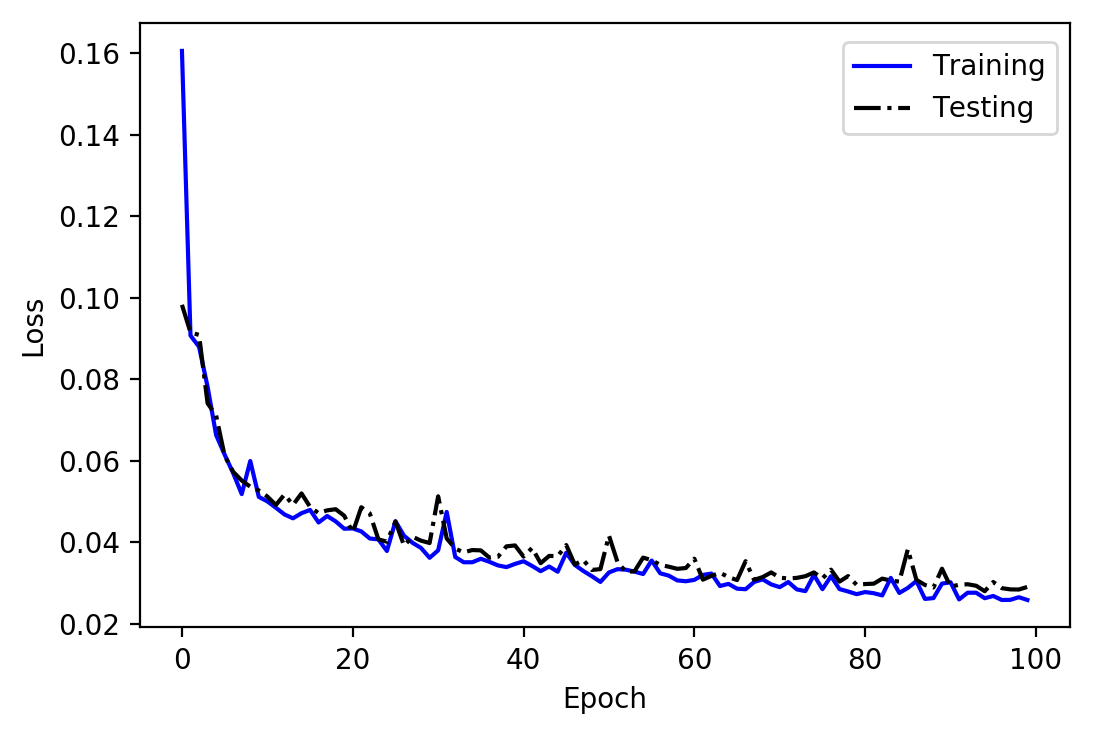}
    \caption{Feedforward neural network training and testing losses}
    \label{fig:loss}
\end{figure}

After conducting hyperparameter grid search for the FNN surrogate model, the optimal architecture of the neural network is listed in Table \ref{tab:fnn_params}. The FNN features 6 layers, ReLU activation, an initial learning rate of 9.5E-04 decayed by a factor of 0.95 following 5 epochs of no improvement, a mini-batch size of 8, and a test split of 0.2. The training and testing losses for the FNN are plotted in Figure \ref{fig:loss}, showing very good convergence trend and negligible overfitting (i.e. training and testing losses are close). 

The validation metrics of the FNN surrogate based on the test set are: MAE=1.59E-5 (15.9 $\mu\epsilon$), RMSE = 3.19E-5 (31.9 $\mu\epsilon$), and $R^2$ = 0.873, showing a good performance over the test simulations. The MAE shows that the FNN model can mis-predict the strain data by only 16$\mu\epsilon$ on average. The $R^2$ value indicates that 87\% of the strain variance is captured by the FNN surrogate. It is worth highlighting here that in the context of optimization, the surrogate does not need to be very accurate as we are not using it for real-time prediction, but rather as a tool to accelerate the optimization process. All our optimization results found by the surrogate will be validated by running the real simulation code (Sierra) later in section \ref{sec:valid}. At this point, we can certainly use the current FNN model as a surrogate for the ENC method.    

\begin{table}[htbp]
  \centering
  \small
  \caption{Feedforward neural network optimized hyperparameters for the ENC method}
    \begin{tabular}{ll}
    \toprule
    Item  & Value \\
    \midrule
    Number of layers & 6 \\
    Number of nodes per layer & 242, 191, 128, 108, 94, 72 \\
    Activation function & ReLU \\
    Initial Learning rate & 9.5E-04 \\
    Loss function & MAE \\
    Batch size & 8 \\
    Number of epochs & 100 \\
    Training data & 482 \\
    Test data & 121 \\
    Test Split & 0.2 \\
    \bottomrule
    \end{tabular}%
  \label{tab:fnn_params}%
\end{table}%

The ENC optimization is carried by running a total of 100 rounds, where in each round an optimizer is randomly selected from the ensemble: ES, MFO, GWO, and DE. Each algorithm is executed for 100 generations with a population size $pop\_size=40$ individuals. This makes the total number of function evaluations (i.e. FNN evaluations) about $100\times100\times4$=400,000, which should be sufficient to explore the search space of the three parameters. The cost of each round is between 2-5 min depending on the algorithm being executed, where the 100 rounds can be executed in parallel due to their independence. 

\begin{figure}[!h]
    \centering
    \includegraphics[width=.65\textwidth]{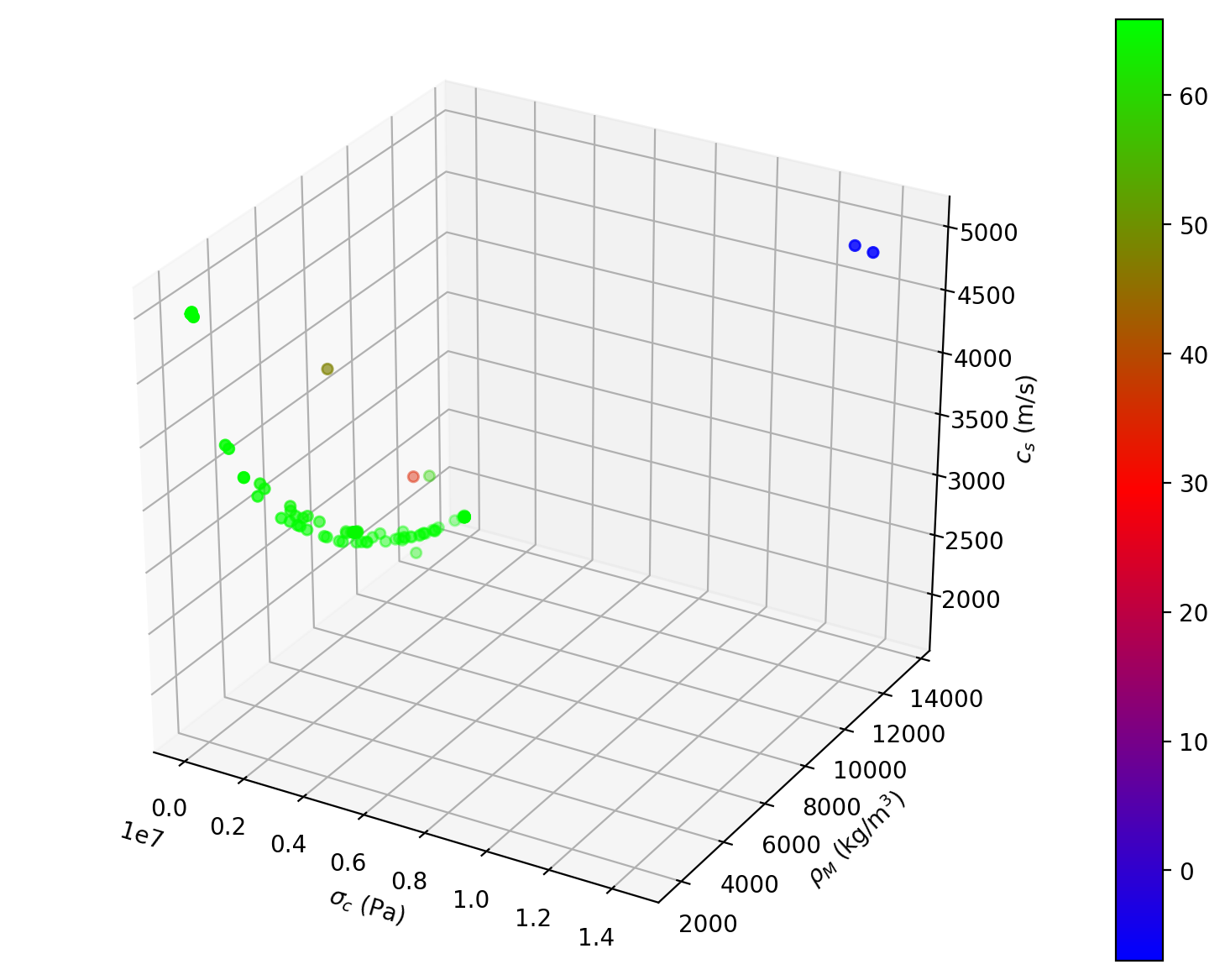}
    \caption{Plot of the best calibrated parameters by ENC and their corresponding objective value (colorbar) for all 100 optimization rounds. Higher color bar values are better}
    \label{fig:jetflow_fnn_opt_grid}
\end{figure}

In Figure \ref{fig:jetflow_fnn_opt_grid}, the best parameter value of each optimization round is plotted in a 4D plot along with the best objective value as estimated by Eq.\eqref{eq:obj}. The objective value is given in the color bar, and since this is a maximization problem, higher objective values are better. We can notice an excellent agreement between the optimization rounds, as most managed to reach the green area that indicates a maximum objective function. Interestingly, the results are indicating that the optimal parameter values for $\vec{x} =[\sigma_c, \rho_M, c_s]$ are making a green curve in Figure \ref{fig:jetflow_fnn_opt_grid}. The results are showing that lower tensile cutoff threshold is preferred, while the density and speed of sound are inversely correlated; for a higher mercury density, the speed of sound is lower. 

\begin{figure}[!h]
    \centering
    \includegraphics[width=.5\textwidth]{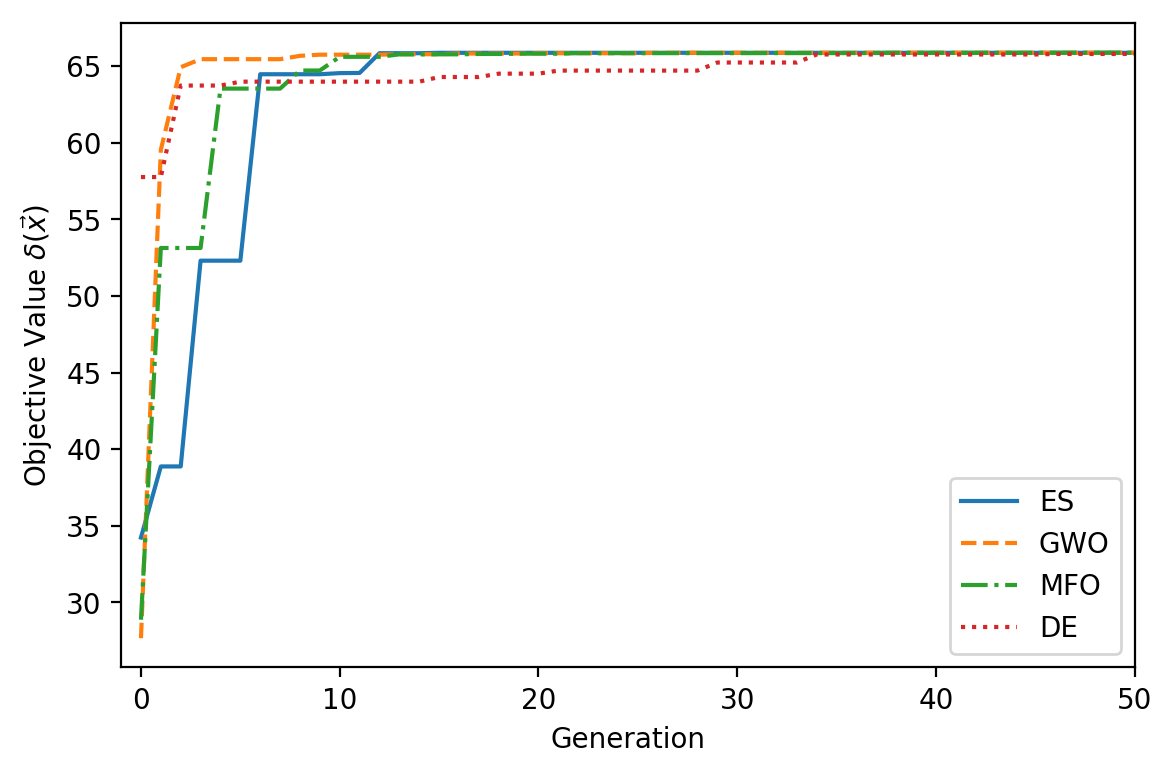}
    \caption{Convergence of the ENC optimization objective function for different algorithms. Only the best round is plotted. For example, if 25 independent ES rounds are executed, each with 100 generations, the round that found the best solution is plotted.}
    \label{fig:jetflow_fnn_opt_conv}
\end{figure}

Also, to confirm the convergence of the four optimization algorithms, Figure \ref{fig:jetflow_fnn_opt_conv} illustrates the convergence curve of the ``best'' round of each optimizer for only 50 generations of the total 100 (i.e. sufficient to show convergence and also differences). It is clearly shown that despite some optimizers being slower at the beginning of the search, all managed to converge to $\delta(\vec{x}) > 65.0$ within 50 generations of search. Nevertheless, not all algorithms are showing consistency across all rounds. Table \ref{tab:opt_stat} demonstrates the mean and standard deviation of the best objective value achieved by each optimizer during the full search process. ES shows a lower mean and a much larger standard deviation, implying that ES was struggling in some rounds to obtain a competitive objective value. In fact, the two blue points in Figure \ref{fig:jetflow_fnn_opt_grid}, which have negative objective value were found by the ES optimizer. GWO, MFO, and DE seem to have decent performance, with DE showing the best and most consistent performance among all. 

\begin{table}[htbp]
  \centering
  \caption{Summary statistics for the best objective value found by each optimizer}
    \begin{tabular}{lll}
    \toprule
    Algorithm & Mean  & Standard Deviation \\
    \midrule
    ES    & 57.89 & 19.87 \\
    MFO   & 65.70 & 0.31 \\
    DE    & 65.73 & 0.24 \\
    GWO   & 65.53 & 0.44 \\
    \bottomrule
    \end{tabular}%
  \label{tab:opt_stat}%
\end{table}%

\subsection{Sparse Polynomial Expansion Results}
\label{sec:SPE_results}
Our investigation reveals that there are many polynomial surrogates that can fit the data with competing accuracy, and those surrogates may indicate different locations for the optimal parameters. For an objective and robust model calibration, instead of relying on a single polynomial surrogate, we investigate a large number of surrogates, which are generated with a wide variety of settings, to form a distribution of candidate parameters. Then, the best parameters will be informed by analyzing the statistics of this distribution. 

We consider $\{\Psi_j\}_{j\in \mathcal{J}}$ to be the Total Degree polynomial spaces with the orders ranging from $10$ to $18$, and Tensor Product spaces with the orders ranging from $5$ to $11$. The degrees of freedom vary from $286$ to $1330$ for the Total Degree spaces, and from $216$ to $1728$ for the Tensor Product spaces. Since we have $480$ training samples (randomly selected from the set of 603 simulations), this range offers reasonable ratios of the number of samples to unknowns, which are commonly used in sparse polynomial recovery. We also construct the surrogates with different values of regularization parameters $\lambda$, see Eq.\eqref{intro:l1}. This parameter handles the risk of overfitting and encourages learning sparse surrogates. It is worth noting that for the majority of samples, the discrepancy values are large, indicating an unpromising search region. It is more beneficial to have simpler surrogates that are able to capture the trend of the discrepancy there, rather than more complicated surrogates that perfectly fit the discrepancy. Towards that end, we investigate multiple $\lambda$ with moderate and large values over a wide range, in particular, $\lambda = 100,\, 500,\, 1000,\, 1500,\, 2000$. The total number of different polynomial surrogates constructed for this study therefore is $(7+9)\times 5 =80$. The training errors of these surrogates in relative RMSE norm ranges from $1.30\%$ to $13.20\%$ with an accumulated mean $5.10\%$ (corresponding to $R^2$ score $0.98$), while the validation errors range from $5.70\%$ to $14.00\%$ with an accumulated mean $8.9\%$ (corresponding to $R^2$ score $0.97$). 

\begin{figure}[!h]
    \centering
    \includegraphics[width=.35\textwidth]{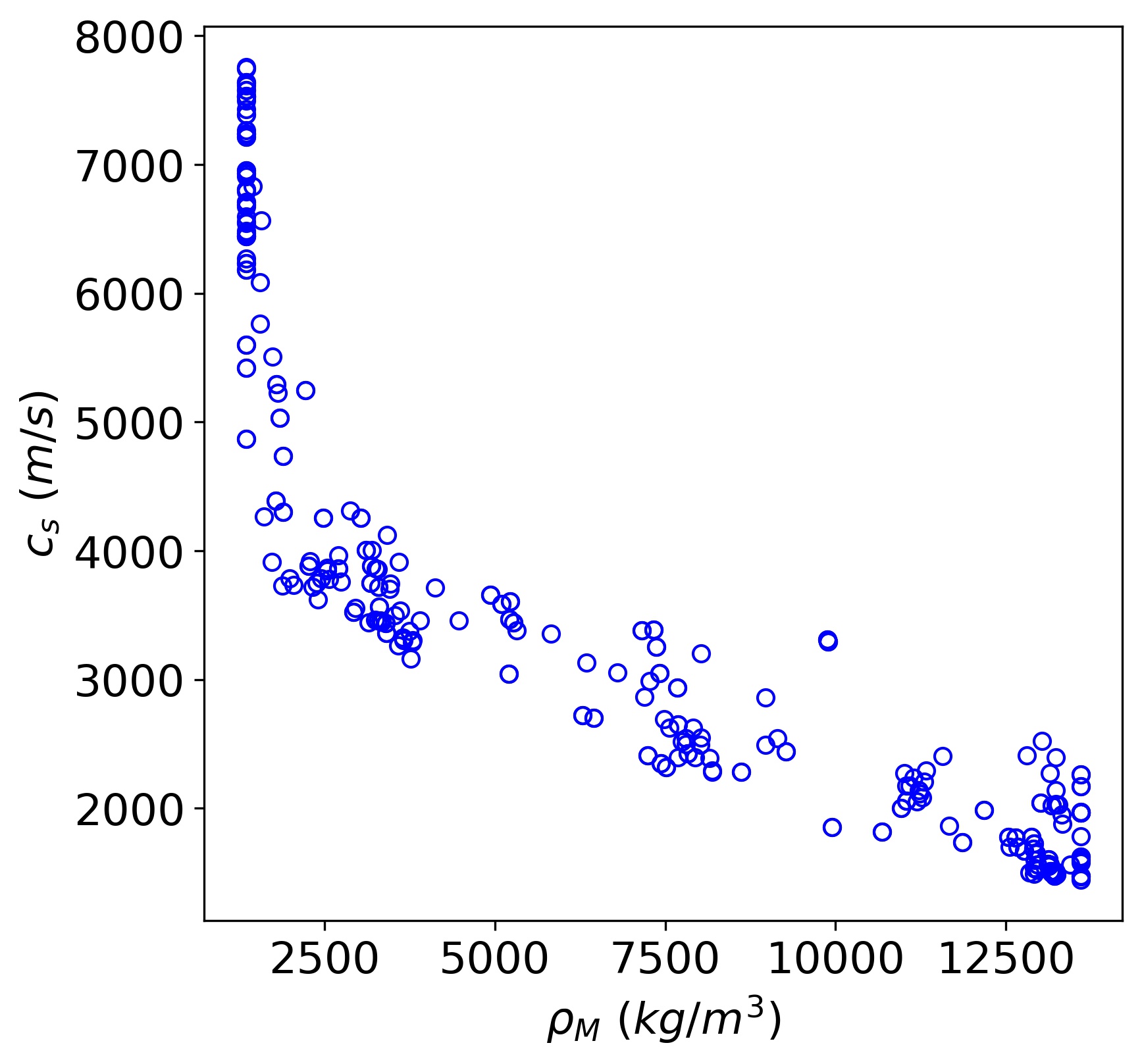}
    \includegraphics[width=.35\textwidth]{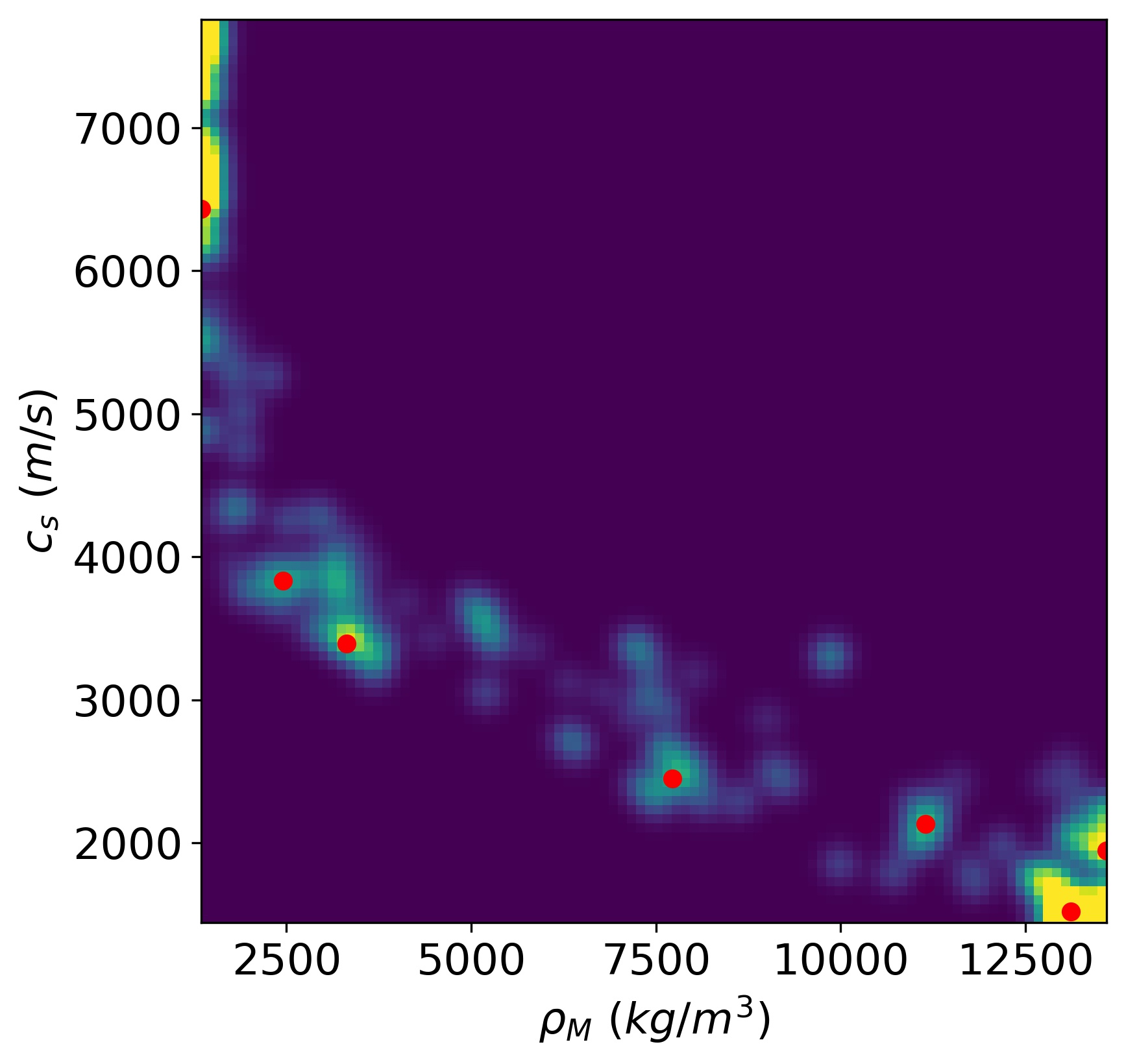}
    \caption{(Left) Distribution of the candidate parameters informed by all 80 investigated polynomial surrogates, identified by DGS algorithm. All tensile cutoff parameters are located close to their lower threshold, therefore, the tensile cutoff dimension is omitted here for better visualization; (Right) The approximating continuous distribution of the candidate parameters. This distribution is multimodal, and we extract the best of candidate parameters by selecting the peak of each mode.}
    \label{fig:SPE_optima}
\end{figure}

For each surrogate, we run 50 trials of DGS algorithm (Algorithm {\color{blue} 2}) and collect the optima identified by the optimization. About 2-5 local optima are found by the run for each polynomial surrogate, bringing the total number of candidate parameters to 309. A run of DGS has 100 iterations, each of which uses 6 function calls for computing the searching directions, and 20 calls for line search, bringing the number of function evaluations to 2600. Thus, we use $2600\times 50 \text{ trials }= 130,000$ evaluations to investigate one surrogate. These parameters concentrate along a curve showing inverse correlation between the density and speed of sound, while the tensile cutoff is always very close to its lower threshold. We plot the locations of candidate parameters projected on the density-speed of sound plane in Figure \ref{fig:SPE_optima} (left). We note that the distribution of parameters here agrees with the convergence curve of ENC approach in Figure \ref{fig:jetflow_fnn_opt_grid}, however, with greater variation, because a large number of surrogate models are involved. We extract the best calibrated parameters from this distribution by selecting the ones that most frequently appear. The distribution is multimodal, where the parameters visibly form several clusters. We apply a Gaussian convolution to find an approximating continuous distribution and select the peak of each mode for optimal parameters. The validation of calibrated parameters is presented in Section \ref{sec:valid}.    

\begin{figure}[!h]
    \centering
    \includegraphics[width=.8\textwidth]{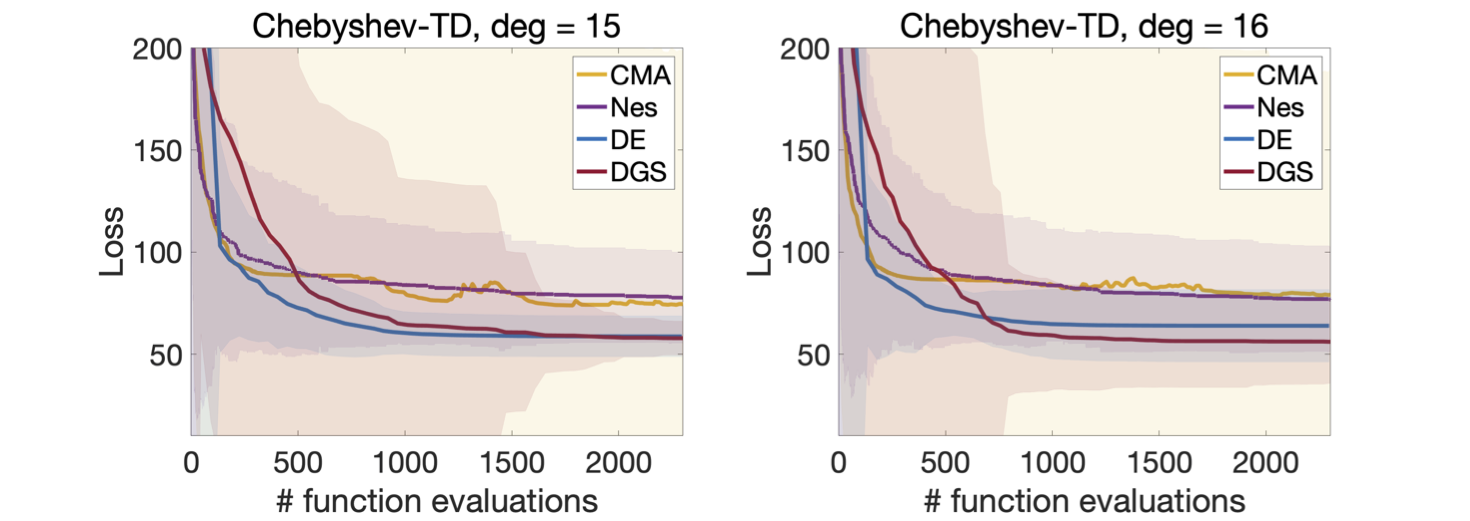}
    \caption{Comparison of the objective function decay with respect to the number of function evaluations for two Chebyshev surrogates on Total Degree subspaces of order 15 and 16. Each curve is the mean of $50$ independent trials and the shaded areas represent [mean-3std, mean+3std]. DGS method achieves the lowest mean loss values in both cases. }
    \label{fig:SPEsurro_error}
\end{figure}

To illustrate the performance of our DGS algorithm, we compare the DGS method with the following derivative-free optimization: (i) {IPop-CMA}: the restart covariance matrix adaptation evolution strategy with increased population size \cite{1554902}. We use the code pycma v3.0.3 available at \texttt{https://github.com/CMA-ES/pycma}; (ii) {Nesterov}: the random search method in \citep{NesterovSpokoiny15}; and (iii) {DE}: differential evolution \citep{storn1997differential}, where we use the implementation from 
\texttt{scipy.optimize.differential\_evolution} and its default hyperparameters. We show the comparative performance of the DGS method on two different surrogates, constructed on Chebyshev Total Degree subspaces of order 15 and 16 with regularization parameter $\lambda=1000$. For each method, we run 50 independent trials with random initialization in the searching domain and plot the decay of objective function with respect to the number of function evaluations. The comparison between DGS and the baselines is shown in Figure \ref{fig:SPEsurro_error}. We find that DGS and DE are more competitive than CMA and random search in this test. Due to the relatively large budget for line search in each iteration, the function decay for DGS is slow at the beginning. However, with nonlocal DGS gradient for long-range exploration, the method is less susceptible to bad local minima and reliably discovers the best minina eventually. DE achieves the same mean objective value in one test and slightly worse in another; however, this method drops the function values faster at the beginning. CMA and Nesterov's random search converge to similar mean objective values, but the variation is much larger for CMA. This indicates that CMA generally is able to find competitive minima, but a small percentage of runs struggling at bad minima hampers the overall performance of the method.   

\subsection{Methodology Validation}
\label{sec:valid}

\begin{figure}[!h]
    \centering
    \includegraphics[width=\textwidth]{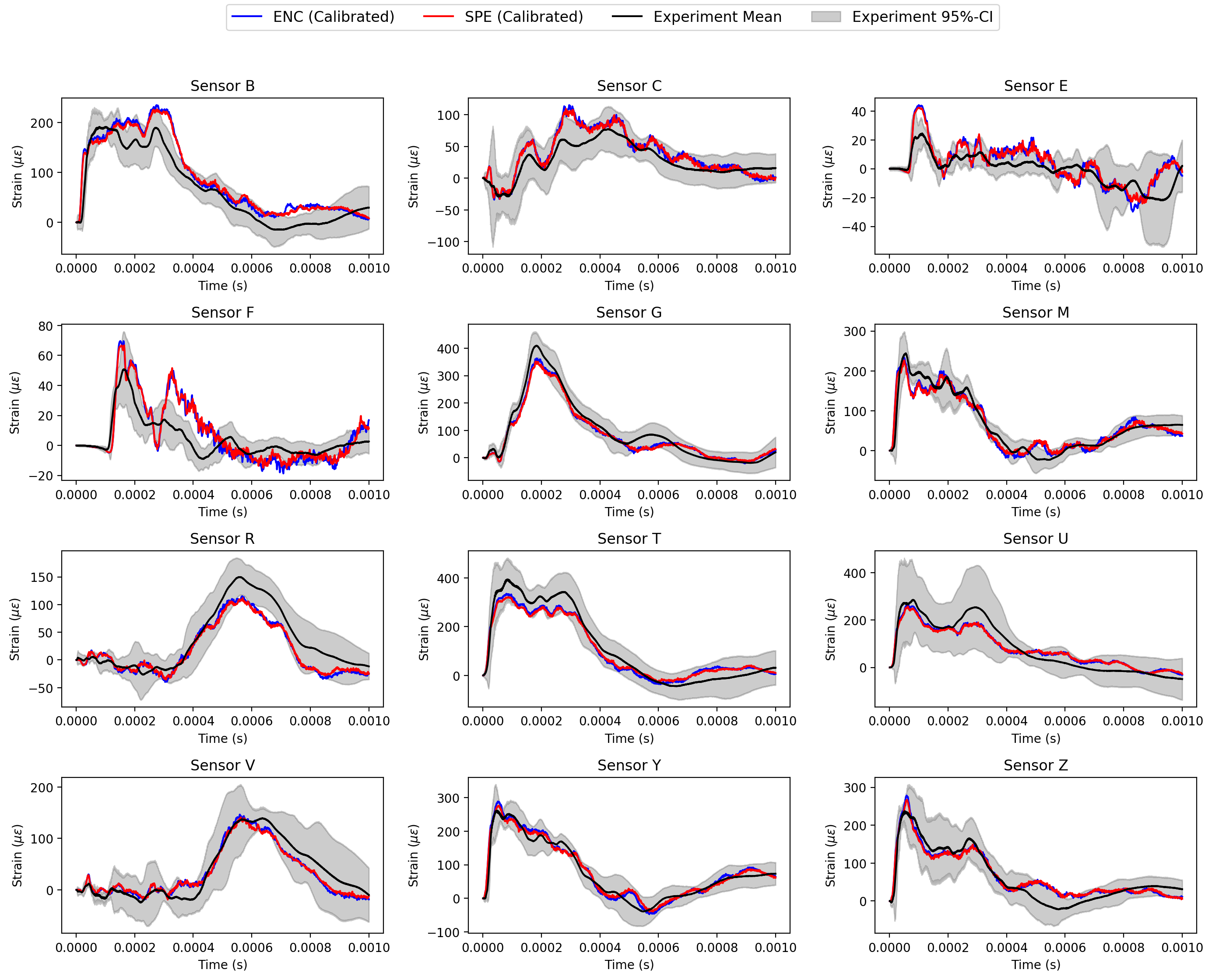}
    \caption{The results of the calibrated simulation against the sensor experimental data using the ENC and SPE methods. The calibrated simulation is executed using the parameters of Table \ref{tab:param_best}.}
    \label{fig:calib}
\end{figure}

The top 5-10 candidates found by ENC and SPE are passed to the computer code Sierra for testing to eliminate the error of the surrogate models. The best calibrated parameters obtained by ENC and SPE are reported in Table \ref{tab:param_best} along with a prior set of parameters recommended by Reimer \cite{riemer2005benchmarking} based upon nominal conditions. The first look at the parameter sets shows that both SPE and ENC recommend significantly lower tensile cutoff threshold. Furthermore, ENC and SPE utilize the inverse relationship between mercury density and speed of sound reflected by the green curve in Figure \ref{fig:jetflow_fnn_opt_grid} and the blue curve of Figure \ref{fig:SPE_optima} (left), showing lower density and higher speed of sound as the optimal parameters for better agreement between the simulation and the experiment. This can be confirmed using Figure \ref{fig:calib} which shows how ENC and SPE-based simulations are compared to the experiment mean and 95\%-CI. First observation is the near-perfect agreement between SPE and ENC despite having different parameter sets. Second observation is the good agreement achieved for most sensors compared to the experimental values, for instance, sensor G, M, R, T, U, V, Y, and Z seem to show a good agreement, while the remaining sensors show a fair agreement. Overall, ENC seems to be closer to the nominal conditions of Riemer \cite{riemer2005benchmarking} than SPE, and both seem to yield good and consistent results. 

An interesting conclusion of this work is that we noticed the best mercury density and sound speed that fit the simulation and experiment are far from the nominal mercury physical properties \cite{riemer2005benchmarking}, especially for SPE, which make their physical justification difficult. There are few reasons that explain this large offset. First, the model-form uncertainty that highlights the limitations of the EOS model in capturing mercury cavitation and multi-phase phenomena of the target. Second reason is the biases and errors in the strain measurements, which are reflected in Figure \ref{fig:calib}. Consequently, the physical model parameters try to compensate for these discrepancy sources to improve simulation fitness to the data. Therefore, we should emphasize here that those settings should be only used for the problem, computer model, and strain data used in this study, and cannot be generalized for other liquid mercury applications. And we can only say these EOS parameters that do not look physically possible generate in fact a better EOS for cavitating mercury in the jet-flow target at 1.4 MW of the spallation neutron source, accounting for model and data uncertainties. As this work focuses mostly on outlining and testing calibration methods, our future work is to expand the current methods to a more challenging and expensive two-phase flow model of Rayleigh-Plesset, at which bubble dynamics are modeled in more detailed, resolving the physical shortcoming of the current EOS model.  

\begin{table}[h]
  \centering
  \small
  \caption{Summary of the best calibrated parameters found by ENC and SPE}
    \begin{tabular}{llll}
    \toprule
    Parameter & ENC   & SPE   & Riemer \cite{riemer2005benchmarking} \\
    \midrule
    $\sigma_c$ & 85642 & 53300 & 150000 \\
    $\rho_M$ & 10345 & 7720  & 13500 \\
    $c_s$ & 2020  & 2449  & 1450 \\
    \bottomrule
    \end{tabular}%
  \label{tab:param_best}%
\end{table}%

\begin{table}[htbp]
  \centering
  \small
  \caption{Comparison of sensor \textbf{accuracy (\%)} by ENC, SPE, and reference against experimental data using the parameters of Table \ref{tab:param_best})}
  \begin{threeparttable}
    \begin{tabular}{llll}
    \toprule
    Sensor & ENC   & SPE   & Riemer \cite{riemer2005benchmarking} \\
    \midrule
    B     & 66    & 69    & 70 \\
    C     & \textbf{92}* & \textbf{93} & 63 \\
    E     & \textbf{67} & \textbf{69} & 64 \\
    F     & \textbf{54} & \textbf{60} & 40 \\
    G     & \textbf{87} & 83    & 85 \\
    M     & \textbf{82} & \textbf{85} & 75 \\
    R     & 80    & 81    & 85 \\
    T     & 99    & 93    & 100 \\
    U     & \textbf{100} & \textbf{100} & 100 \\
    V     & \textbf{92} & 90    & 91 \\
    Y     & \textbf{96} & \textbf{97} & 70 \\
    Z     & \textbf{88} & \textbf{86} & 83 \\
    Average & \textbf{84} & \textbf{84} & 77 \\
    \bottomrule
    \end{tabular}%
    \begin{tablenotes}
      \small
      \item $*$ Bold text represents sensors that exhibit similar or higher accuracy than Riemer proposed values \cite{riemer2005benchmarking} 
    \end{tablenotes}
    \end{threeparttable}
  \label{tab:acc}%
\end{table}%

For a quantitative analysis of the performance of the new calibrated parameters, we report two universal metrics: accuracy in Table \ref{tab:acc} based on Eq.\eqref{eq:f2} and mean absolute error ($\mu \epsilon$) in Table \ref{tab:error} based on Eq.\eqref{eq:f1}. For the accuracy metric, which represents the fraction of the simulation points that fall within the confidence bounds, both SPE and ENC methods retain similar average accuracy of 84\%, with 7\% higher than the accuracy obtained by the Riemer reference parameters \cite{riemer2005benchmarking}. In addition, we can notice that ENC was able to improve or maintain the accuracy of 9 out of 12 sensors compared to the reference parameters \cite{riemer2005benchmarking}, while SPE managed to do so for 7 sensors. Some sensors experienced a magnificent improvement such as sensor C, Y, and F, while some other sensors experienced a marginal improvement such as sensor V and G. We can notice that both methods tend to sacrifice some sensors to make a significant improvement into the others such as sensor B, R, and T, thankfully, the reduction is small (within 5\% accuracy for ENC and 7\% for SPE). Similarly, we can see a small reduction in MAE compared to the experiment mean as in Table \ref{tab:error}, and a very good agreement between ENC and SPE's average MAE. 

In this study, the authors intentionally proposed two different formulations of the objective function in Eq.\eqref{eq:obj} and Eq.\eqref{eq:loss}, which were provided to the optimizer. While both formulations have the same goal (i.e. minimize the difference between simulation and experiment mean and maximize number of simulation points in the experiment 95\% CI), the two functions are expressed in different ways mathematically. This again reveals that this part of both ENC and SPE is problem and user-dependent, and typically different formulations should be tested before finding the most optimal form. In addition, the difference in the nature of the optimization algorithms used in this work reflects also additional flexibility. For ENC, we used population-based gradient-free evolutionary algorithms (e.g. ES, DE, GWO, MFO), while for SPE, we utilized a first order scheme equipped with nonlocal DGS gradients, and both categories yielded competitive performance and very good agreement in the results. Both methods also share the concept of diversifying the search process to achieve a good coverage of the search space. ENC does that by running different rounds of different optimization algorithms with different random seed (i.e. initial guess), while SPE employs a large number of polynomial surrogates, constructed on different polynomial subspaces and with different regularization parameter $\lambda$. 

\begin{table}[htbp]
  \centering
  \small
  \caption{Comparison of sensor \textbf{mean absolute error ($\mu\epsilon$)} by ENC, SPE, and reference against experimental data using the parameters of Table \ref{tab:param_best}}
    \begin{tabular}{llll}
    \toprule
    Sensor & ENC   & SPE   & Riemer \\
    \midrule
    B     & 26.6  & 26.4  & 26.6 \\
    C     & 14.7  & 13.2  & 23.8 \\
    E     & 7.6   & 7.1   & 8.1 \\
    F     & 9.8   & 9.1   & 13.1 \\
    G     & 18.9  & 20.5  & 20.1 \\
    M     & 16.0  & 14.9  & 20.9 \\
    R     & 20.6  & 20.1  & 17.6 \\
    T     & 32.7  & 33.6  & 24.7 \\
    U     & 29.4  & 31.6  & 25.9 \\
    V     & 14.6  & 14.4  & 13.9 \\
    Y     & 14.1  & 13.1  & 28.1 \\
    Z     & 18.1  & 18.7  & 21.5 \\
    Average & 18.6  & 18.6  & 20.4 \\
    \bottomrule
    \end{tabular}%
  \label{tab:error}%
\end{table}%

Although both ENC and SPE seem to share multiple features in common and have good agreement in terms of how the simulation is compared to the experiment, each approach has its own advantages and disadvantages. For ENC, the approach features more flexibility when it comes to the objective function. Since the surrogate is used for the computer model, changing the objective function formulation in Eq.\eqref{eq:obj} results in no change in the surrogate structure. Second, ENC is expected to scale well with problem dimensionality due to the neural network modeling capabilities. In this study $n_x=3$, and for $n_x > 20$, ENC is expected to maintain good performance if sufficient data is generated. On the other hand, ENC suffers from the data greediness of neural networks as well as computing costs of network training. Also, the fact that ENC surrogate is predicting a 2D array of time-dependent strain in all sensors is a challenge for the ENC surrogate accuracy. These issues are resolved by the SPE method. SPE method allows more data-economic polynomial-based sparse surrogate fitting with shorter computing time. In addition, since the SPE surrogate only predicts a scalar discrepancy value, it is much less challenging than ENC, which explains why $R^2 > 0.95$ for SPE and $0.87$ for ENC. However, SPE is dependent on the objective/discrepancy function provided to the surrogate, e.g. Eq.\eqref{eq:loss}. When $n_x$ grows, the size of the polynomial subspaces can quickly expand, and it is challenging for polynomial surrogates to capture objective functions with strong local structures, steep edges, or optima with small attraction areas. In those cases, we must have a more refined construction of the polynomial subspaces, e.g., with localized basis and focusing on promising areas, to maintain a reasonable size of the subspaces and data efficiency. Accordingly, we typically recommend choosing the appropriate approach depending on the problem at hand or using both for a comprehensive analysis. 

Despite the good results and improvement achieved in this work, there are two main limitations for both ENC and SPE that should be highlighted: (1) the ``point estimate'' assumption of the model parameters and (2) inherently capturing the hidden correlation between the parameters as was observed in the green curve in Figure \ref{fig:jetflow_fnn_opt_grid}. Both of these limitations are expected to be resolved by the usage of the probabilistic Bayesian inference framework that assumes a probability distribution over the model parameters, and quantifies marginal and joint density distributions by including the correlation between the model parameters. Nevertheless, the Bayesian inference is known to be more complex mathematically and computationally intensive (mainly due to the Markov chain Monte Carlo sampling) as our output space is huge (10,000 time steps for 12 sensors). This is where our proposed ENC and SPE methods excel by providing more flexible, easy-to-implement, and fast calibration methods. Therefore, our future study will focus on resolving the hurdles that face the application of Bayesian inference to this problem to validate the ``frequentist'' approach adopted in this study. 

The value of these calibration efforts to the SNS mission is significant. The strain/stress results from finite element simulation are utilized in fatigue analysis to estimate component’s lifetime and integrity. Thus, the strain response collected from simulation, particularly the strain peaks and the strain range of the profile are critical in estimating steel vessel’s fatigue cycles. Unfortunately, fatigue diagnostics using the strain sensor measurements are not possible due to the sensor failure due to radiation damage after the first few days of the target operation for most of the front sensors. More critically, we are currently unable to place sensors at the highest stress locations on the inside of the mercury target. Therefore, an accurate simulation of the target physics is crucial for accurate fatigue analysis. In addition, our analysis shows the predicted strain peaks by ENC/SPE for the sensors with lower accuracy (e.g. B, E, F) are higher than the experimental mean values during the early pulse time (see Figure \ref{fig:calib}). These over-predicted strain peaks will lead to shorter prediction of target vessel’s lifetime, which tends to be more conservative in target design; biasing designs toward more reliable operation.

The  simulations for the response of the target to a pulse require intensive computational resources and are not practical for searching large parameter spaces. Therefore, our computational algorithms featuring fast surrogate models to tie the experimental data to the physical parameters are a natural approach to expedite the development of a highly fidelity model that can be used to advance target designs in the future. 

\section{Conclusions}
\label{sec:conc}

In this work, we have presented two methods based on neural networks and sparse polynomial expansion surrogate modeling to perform model calibration of computer simulations. The first method is called evolutionary neural calibration, which employs feedforward neural network surrogate model of the solid mechanics simulation. Then the surrogate model is coupled with evolutionary/swarm optimization algorithms and measured data to minimize the discrepancy between the surrogate model and the measured data. The second approach is a classical sparse polynomial approximation trained to directly minimize the discrepancy between the simulations and measurements. The two approaches show competitive results and very good agreement for the same problem. The two methods were applied to calibrate the simulations of the mercury target in the spallation neutron source facility. The calibration methods were able to find optimal simulation settings for the tensile cutoff threshold, mercury density, and mercury speed of sound that reduce the strain discrepancy between simulations and sensor measurements, and thereby improving existing literature results in this area. The newly calibrated simulations achieve 7\% average improvement on sensor accuracy and 8\% reduction in mean absolute error compared to previously reported reference parameters. Some individual sensors experienced up to 30\% improvement in their accuracy using the calibrated parameters. 

The generic nature of the evolutionary neural calibration and sparse polynomial expansion may imply their applicability to calibration problems in other domains. While this study treats the calibrated parameters as a point estimate, our future study will utilize the Bayesian inversion framework, where a random distribution of the calibrated parameters will be determined using individual experiments rather than a lumped mean and confidence interval. In addition, these methods will form the basis to tackle the more advanced target simulations where helium gas is injected into the mercury target to reduce strain and cavitation damage. Given the limitations of the current constitutive model of the equation of state, two-phase models that capture bubble dynamics will be pursued, however, with increasing challenges from both computational and physics perspectives.  

\section*{Acknowledgment}

The authors are grateful for support from the Neutron Sciences Directorate at ORNL in the investigation of this work. This work was supported by the DOE Office of Science under grant DE-SC0009915 (Office of Basic Energy Sciences, Scientific User Facilities program). A portion of this research used resources at the Spallation Neutron Source, a DOE Office of Science User Facility operated by the Oak Ridge National Laboratory. This research used resources of the Computer and Data Environment for Science (CADES) at the Oak Ridge National Laboratory, which is supported by the Office of Science of the U.S. Department of Energy under Contract No. DE-AC05-00OR22725. The authors have also used resources of the Argonne Leadership Computing Facility, which is a DOE Office of Science User Facility supported under Contract DE-AC02-06CH11357.

Notice: This manuscript has been authored by UT-Battelle, LLC, under contract DE-AC05-00OR22725 with the US Department of Energy (DOE). The US government retains and the publisher, by accepting the article for publication, acknowledges that the US government retains a nonexclusive, paid-up, irrevocable, worldwide license to publish or reproduce the published form of this manuscript, or allow others to do so, for US government purposes. DOE will provide public access to these results of federally sponsored research in accordance with the DOE Public Access Plan (http://energy.gov/downloads/doe-public-access-plan).

\section*{CRediT Author Statement}

\noindent \textbf{Majdi I. Radaideh}: Conceptualization, Methodology, Software, Validation, Investigation, Data curation, Visualisation, Formal analysis, Writing - Original Draft. \\
\textbf{Hoang Tran}: Conceptualization, Methodology, Software, Validation, Formal analysis, Visualisation, Writing - Original Draft. \\
\textbf{Lianshan Lin}: Conceptualization, Methodology, Software, Validation, Writing - Original Draft. \\
\textbf{Hao Jiang}: Conceptualization, Data curation, Writing – Review and Edit. \\
\textbf{Drew Winder}: Conceptualization, Methodology, Project Administration, Writing – Review and Edit. \\
\textbf{Sarma Gorti}: Conceptualization, Investigation, Writing – Review and Edit. \\
\textbf{Guannan Zhang}: Conceptualization, Methodology, Writing – Review and Edit. \\
\textbf{Justin Mach}: Conceptualization, Methodology, Writing – Review and Edit. \\
\textbf{Sarah Cousineau}: Conceptualization, Funding acquisition, Resources, Writing – Review and Edit.


\bibliographystyle{elsarticle-num}
\setlength{\bibsep}{0pt plus 0.3ex}
{
\footnotesize \bibliography{references}}

\end{document}